\documentclass[aps,nofootinbib,groupaddress,twocolumn,floatfix,letterpaper]{revtex4}
\usepackage{graphicx}
\usepackage{epstopdf}
\usepackage{amsmath,amsthm,amssymb,latexsym,amsfonts,times,mathrsfs,verbatim}
\usepackage{mathptmx}
\usepackage{braket}

\DeclareMathAlphabet{\mathcal}{OMS}{cmsy}{m}{n}

\begin{document}

\title{Dueling Dynamical Backaction in a Cryogenic Optomechanical Cavity}
\author{B.D. Hauer}\email{bhauer@ualberta.ca}
\author{T.J. Clark}
\author{P.H. Kim}
\author{C. Doolin}
\author{J.P. Davis}\email{jdavis@ualberta.ca}

\affiliation{Department of Physics, University of Alberta, Edmonton, Alberta, Canada T6G 2E9}

\date{\today}

\begin{abstract}

Dynamical backaction has proven to be a versatile tool in cavity optomechanics, allowing for precise manipulation of a mechanical resonator's motion using confined optical photons. In this work, we present measurements of a silicon whispering-gallery-mode optomechanical cavity where backaction originates from opposing radiation pressure and photothermal forces, with the former dictating the optomechanical spring effect and the latter governing the optomechanical damping. At high enough optical input powers, we show that the photothermal force drives the mechanical resonator into self-oscillations for a pump beam detuned to the lower-frequency side of the optical resonance, contrary to what one would expect for a radiation-pressure-dominated optomechanical device. Using a fully nonlinear model, we fit the hysteretic response of the optomechanical cavity to extract its properties, demonstrating that this non-sideband-resolved device exists in a regime where photothermal damping could be used to cool its motion to the quantum ground state.

\end{abstract}

\maketitle

\section{Introduction}
\label{intro}

Over the past several years, the field of cavity optomechanics, which studies the interplay between the photonic modes of an optical cavity and the phononic modes of a mechanical resonator, has proven itself to be a tremendous resource. Along with numerous applications in precision metrology \cite{teufel_2009, krause_2012, gavartin_2012, kim_2016} and quantum information \cite{palomaki_2013a, riedinger_2016, reed_2017}, optomechanical systems have also demonstrated potential in providing an experimental testbed to help answer a number of fundamental questions in quantum mechanics \cite{bose_1999, marshall_2003, romeroisart_2011, chen_2013} and gravity \cite{pikovski_2012, abbott_2016, pfister_2016, belenchia_2016}. One of the most powerful effects observed in optomechanical systems, however, has proven to be the ability of the photons in the cavity to manipulate the mechanical resonator's motion via the radiation pressure force. Using the delayed nature of this interaction, which arises due to the finite lifetime of photons in the cavity, the optical field is able to modify the properties of the mechanical resonator, resulting in dynamical backaction between the optical and mechanical modes of the system \cite{aspelmeyer_2014}. 

By detuning an optical pump to the lower frequency (red) side of the optical cavity, energy can be extracted from the mechanical resonator via scattering processes that preferentially promote detuned photons into the higher energy states of the optical cavity. The opposite process will then occur for a pump detuned to the higher frequency (blue) side of the cavity, such that the photons instead provide energy to the mechanical element, thus amplifying its motion. Accompanying each of these dynamical backaction processes is an increase in the mechanical damping rate, or cooling of the resonator's motion, for a red-detuned optical pump, while for a blue-detuned pump, the damping rate decreases. This process, known as optomechanical damping, is mirrored by the optomechanical spring effect, which results in a decrease (increase) in the resonance frequency of the mechanical oscillator for a red (blue) detuned pump, such that these two dynamical backaction effects obey the Kramers-Kronig relations \cite{aspelmeyer_2014b}. Employing these radiation-pressure-driven effects, a number of groundbreaking experiments have been performed using optomechanical cavities, including motional ground state cooling of micro/nanomechanical resonators \cite{teufel_2011, chan_2011}, entanglement of photonic and phononic modes \cite{palomaki_2013b, riedinger_2018, ockeloen-korppi_2018} and preparation of other nonclassical states of mechanical motion \cite{wollman_2015, pirkkalainen_2015, riedinger_2016}.

Though efforts have largely focussed on this radiation-pressure-driven interaction, optomechanical coupling can be mediated by other means, such as the photothermal (or bolometric) force, whereby photon absorption in the mechanical element introduces a temperature gradient across the device, causing it to deflect due to differential thermal contractions \cite{pinard_2008, deliberato_2011, restrepo_2011, abdi_2012a, abdi_2012b}. Photothermal effects have historically been studied in optical cavities comprised of gold-plated cantilevers \cite{mertz_1993, metzger_2004, metzger_2008a, metzger_2008b, jourdan_2008}, but have also been observed in buckled microcavities \cite{aubin_2004, yuvaraj_2012}, multilayered Bragg mirror beams \cite{gigan_2006}, thin metallic mirrors \cite{zaitsev_2011}, membranes \cite{barton_2012, usami_2012}, nanowires \cite{aubin_2004, hosseini_2014, dealba_2017} and superfluid helium \cite{harris_2016, kashkanova_2017a, kashkanova_2017b}. As in the case of radiation-pressure-driven optomechanics, photothermal forces can also be used to manipulate the motion of mechanical resonators. In fact, in a somewhat paradoxical sense, photothermal coupling can in principle be used to cool a resonator's motion to occupancies of less than a single phonon on average \cite{pinard_2008, deliberato_2011, restrepo_2011, abdi_2012a, abdi_2012b}. Furthermore, photothermal dynamical backaction effects are peculiar in that they are able to invert the detuning dependence of the optomechanical damping (and spring effect) with respect to that found in conventional radiation-pressure-driven systems, where such a reversal is only possible for cavities that are externally-driven to large mechanical amplitudes in the sideband-resolved regime \cite{marquardt_2006, krause_2015, buters_2015}. This results in amplification of the resonator's motion (accompanied by an increase in the mechanical resonance frequency) for red-detuned pumps, while cooling (along with a decrease in the mechanical resonance frequency) occurs for blue-detuned pumps \cite{metzger_2004, metzger_2008a, metzger_2008b, jourdan_2008, yuvaraj_2012}, seemingly violating the conservation of energy. 

While there have been brief mentions of a radiation-pressure-dominated spring effect observed in photothermally-driven optomechanical cantilevers \cite{metzger_2004, metzger_2008a, jourdan_2008}, to date there has not been a thorough experimental investigation of how the photothermal and radiation pressure forces interact with each other. Therefore, a comprehensive study of this interaction is warranted, especially in the case of cryogenic silicon optomechanical cavities, as these devices are integral to a number of quantum optomechanical experiments \cite{cohen_2015, riedinger_2016, hong_2017, riedinger_2018, marinkovic_2018}.

In this article, we present and quantitatively analyze measurements of a silicon whispering-gallery-mode optomechanical cavity that exists in a parameter regime where both radiation pressure and photothermal effects are relevant. We begin by providing a brief theoretical overview of a nonlinear optomechanical system that is subject to both of these forces. Applying this theory to the studied device, we find that radiation pressure dominates the optical spring effect, while the photothermal force governs the system's optomechanical damping. Moreover, the photothermal force acts to oppose its radiation pressure counterpart, such that the optomechanical damping has the opposite detuning-dependence from what one would expect for a conventional radiation-pressure-driven system, resulting in an oddly similar detuning-dependence between the optomechanical damping and spring effect. With this photothermal enhancement to the optomechanical damping, we find that for high enough optical input powers we are able to drive the mechanical resonator into self-oscillations using a red-detuned pump. We show that in this self-oscillating regime, the DC transmission through the optical cavity, as well as the optomechanical damping and spring effect, become highly nonlinear, while demonstrating hysteretic behavior depending on the sweep direction of the optical drive. Using our fully nonlinear treatment of the system, we fit these data, extracting the optomechanical properties of the system. From these experimentally-determined system parameters, we assess the device's ability to cool the motion of the resonator using the photothermal effect for a blue-detuned optical pump. In doing so, we find that the mechanical occupancy can in principle be reduced to less than a single phonon on average, despite the fact that the cavity resides deeply in the non-sideband-resolved regime.

\section{Theoretical Model}
\label{theory}

To describe the behavior of the device studied in this work, we consider an optomechanical system comprised of an optical cavity, with resonant frequency $\omega_{\rm c}$ and total loss rate $\kappa$, coupled to a mechanical resonator, with a resonant frequency $\omega_{\rm m}$ and intrinsic damping rate $\Gamma_{\rm m}$. We assume that a dispersive coupling arises between these two systems due to the fact that the displacement $x$ of the mechanical oscillator shifts the resonance frequency of the optical cavity by an amount $Gx$, where $G = - d \omega_{\rm c} / dx$ is the optomechanical coupling coefficient. For such a system, the coupled classical equations of motion will be given by \cite{harris_2016}
\begin{gather}
\label{aeom}
\dot{a} = -\frac{\kappa}{2} a + i \Delta_0 a + i G x a + \sqrt{\kappa_{\rm e}} \bar{a}_{\rm in}, \\
\label{xeom}
\ddot{x} + \Gamma_{\rm m} \dot{x} + \omega_{\rm m}^2 x = \frac{1}{m} \left[ F_{\rm th} + F_{\rm rp} + F_{\rm pt} \right].
\end{gather}
Equation \eqref{aeom} describes the time evolution of the optical cavity's field amplitude $a$, which is driven by an input field $\bar{a}_{\rm in}$. This drive field, whose strength is related to the input power $P_{\rm in}$ of the signal via the relation $|\bar{a}_{\rm in}|^2 = P_{\rm in} / \hbar \omega_{\rm d}$, is coupled into the optical cavity at a rate $\kappa_{\rm e}$, with its frequency $\omega_{\rm d}$ detuned from cavity resonance by an amount $\Delta_0 = \omega_{\rm d} - \omega_{\rm c}$. With this definition of detuning, negative (positive) values indicate a red-detuned (blue-detuned) cavity drive. Meanwhile, Eq.~\eqref{xeom} governs the dynamics of the mechanical resonator's displacement, which is simultaneously driven by an intrinsic thermal force $F_{\rm th}$, a radiation pressure force $F_{\rm rp} = \hbar G a^\dag a$ \cite{aspelmeyer_2014, harris_2016}, and a photothermal force \cite{pinard_2008, restrepo_2011, abdi_2012a, abdi_2012b}
\begin{equation}
F_{\rm pt}(t) = \frac{\hbar G \beta}{\tau} \int_{-\infty}^t e^{-\frac{t-t'}{\tau}} a^\dag(t') a(t') dt'.
\label{Fpt}
\end{equation}
Here the dimensionless parameter $\beta$ sets the relative strength of the photothermal force with respect to the radiation pressure force and is heavily dependent on the optical and mechanical modeshapes being considered \cite{jourdan_2008, harris_2016}, as well as the thermal properties of the resonator \cite{abdi_2012a, abdi_2012b}. We further note that $\beta$ can be negative, such that the photothermal force acts to directly oppose radiation pressure effects \cite{restrepo_2011, metzger_2004, metzger_2008a, metzger_2008b, jourdan_2008, yuvaraj_2012}, which has very important consequences for the detuning-dependence of the optomechanical damping and spring effects. Also included in Eq.~\eqref{Fpt} is the thermal relaxation time $\tau$ that sets the timescale of the photothermal force and, similar to $\beta$, is determined by the thermal properties and geometry of the device (see Appendix \ref{tauthsec}). Finally, while we have chosen to explicitly identify Eq.~\eqref{Fpt} as a being photothermal in nature, the following analysis is valid for any optomechanical force that has a delayed response with respect to the occupation of the optical cavity.

In order to solve Eqs.~\eqref{aeom} and \eqref{xeom}, we begin by assuming that for a high-$Q$ mechanical resonator ($Q_{\rm m} = \omega_{\rm m} / \Gamma_{\rm m} \gg 1$) driven to a large amplitude of motion, the mechanical displacement will be well-described by the ansatz $x(t) = \bar{x} + A \cos( \omega_{\rm m} t)$, where $\bar{x}$ and $A$ are the resonator's static displacement and amplitude of oscillation, respectively \cite{marquardt_2006, ludwig_2008, aspelmeyer_2014}. Inputting this ansatz into Eq.~\eqref{aeom}, we find \cite{marquardt_2006, aspelmeyer_2014, krause_2015}
\begin{equation}
\label{aoft}
a(t) = \sqrt{\kappa_{\rm e}} \bar{a}_{\rm in} e^{i \phi(t)} \sum_{k=-\infty}^\infty \alpha_k e^{i k \omega_{\rm m} t},
\end{equation}
with $\phi(t) = \xi \sin (\omega_{\rm m} t)$ the time-dependent global phase of the optical field and 
\begin{equation}
\label{alpha}
\alpha_k = \frac{J_k(-\xi)}{\kappa/2 - i \left(\Delta_0 + G \bar{x} - k \omega_{\rm m} \right)},
\end{equation}
where $J_k(z)$ is the $k$th Bessel function of the first kind and $\xi = G A / \omega_{\rm m}$ is the dimensionless mechanical modulation strength \cite{marquardt_2006, ludwig_2008, krause_2015}. Using these solutions for $x(t)$ and $a(t)$, the mechanical-amplitude-dependent optomechanical spring effect and damping are found to be (see Appendix \ref{omeoms})
\begin{align}
\label{dwmnl}
\delta \omega_{\rm m} = -\frac{\hbar G \kappa_{\rm e} |\bar{a}_{\rm in}|^2}{A m \omega_{\rm m}} \sum_{k=-\infty}^\infty {\rm Re} \left\{ \alpha_k \alpha_{k+1}^* \left( 1 + \frac{\beta}{1 - i \omega_{\rm m} \tau} \right) \right\}, \\
\label{dGmnl}
\delta \Gamma_{\rm m} = \frac{2 \hbar G \kappa_{\rm e} |\bar{a}_{\rm in}|^2}{A m \omega_{\rm m}} \sum_{k=-\infty}^\infty {\rm Im} \left\{ \alpha_k \alpha_{k+1}^* \left( 1 + \frac{\beta}{1 - i \omega_{\rm m} \tau} \right) \right\},
\end{align}
from which we can find the total mechanical damping rate as $\Gamma_{\rm tot} = \Gamma_{\rm m} + \delta \Gamma_{\rm m}$. Finally, using Eq.~\eqref{aoft}, along with the input/output relation $\bar{a}_{\rm out} = \bar{a}_{\rm in} - \sqrt{\kappa_{\rm e}} a$, we can also determine the DC transmission through the optical cavity as \cite{aspelmeyer_2014}
\begin{equation}
\label{Tdcnl}
\mathcal{T}_{\rm DC} = \frac{\bar{a}_{\rm out}}{\bar{a}_{\rm in}} = 1 - 2 \kappa_{\rm e} {\rm Re} \left\{ \sum_{k=-\infty}^\infty J_{-k}(\xi) \alpha_k \right\} + \kappa_{\rm e}^2 \sum_{k=-\infty}^\infty |\alpha_k|^2.
\end{equation}
We note that by taking the small mechanical amplitude limit ({\it i.e.}, $\xi \ll 1$) for each of the quantities in Eqs.~\eqref{dwmnl}, \eqref{dGmnl} and \eqref{Tdcnl}, their standard linearized expressions can be obtained (see Appendix \ref{omeoms}).

\begin{figure}[t!]
\centerline{\includegraphics[width=\columnwidth]{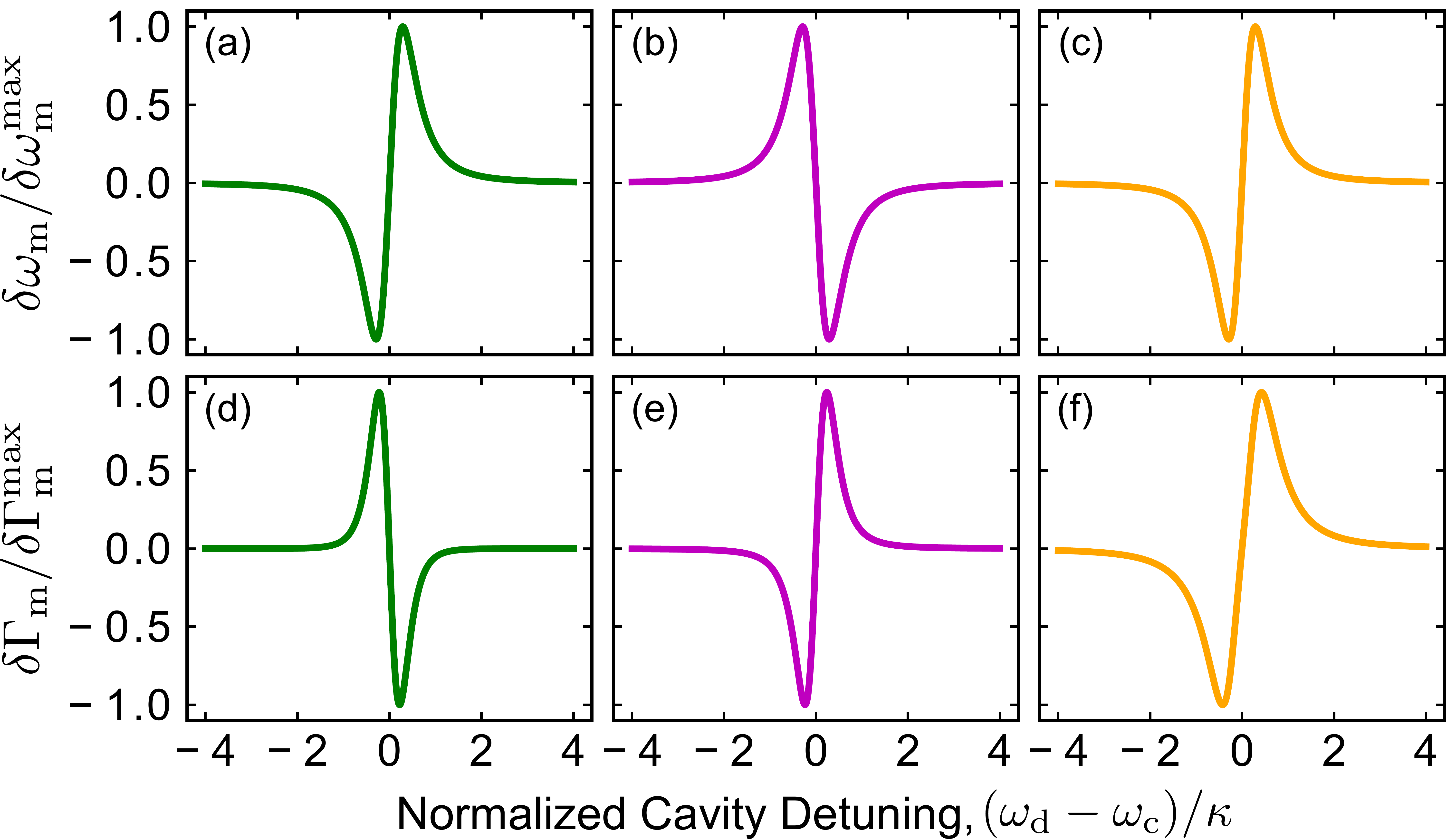}}
\caption{{\label{fig1}} Plots of the optomechanical (a)--(c) spring effect and (d)--(f) damping, normalized to their respective maximum values, $\delta \omega_{\rm m}^{\rm max}$ and $\delta \Gamma_{\rm m}^{\rm max}$. Here we consider the small mechanical amplitude regime with both radiation pressure and photothermal effects included ({\it i.e.}, Eqs.~\eqref{dwmnl} and \eqref{dGmnl} in the limit $\xi \ll 1$ -- see Appendix \ref{omeoms}), where we have taken (a)/(d) $\beta$ = -0.1, $\omega_{\rm m} \tau$ = 0.01, (b)/(e) $\beta$ = -10, $\omega_{\rm m} \tau$ = 0.01, and (c)/(f) $\beta$ = -0.1, $\omega_{\rm m} \tau$ = 1; with $\kappa / \omega_{\rm m}$ = 100 in all plots. Therefore, according to Eqs.~\eqref{dwmptdomfin} and \eqref{dGmptdomfin}, in (a)/(d) radiation pressure dominates both the spring effect and damping, while in (b)/(e) dynamical backaction is driven by photothermal forces. Finally, in (c)/(f) we enter into the dueling regime studied in this work, where the spring effect is dominated by the radiation pressure force, while the optomechanical damping is governed by photothermal effects.}
\end{figure}

The two expressions in Eqs.~\eqref{dwmnl} and \eqref{dGmnl} resemble what one would expect for a radiation-pressure-driven optomechanical system that has been driven to large mechanical amplitude, with the addition of the $\beta / (1 - i \omega_{\rm m} \tau)$ term to account for the photothermal force. Photothermal forces will therefore govern these dynamical backaction effects when this additional term overwhelms its radiation pressure counterpart (see Appendix \ref{omeoms}). For a non-sideband-resolved optomechanical cavity, this occurs for the spring effect when 
\begin{equation}
\label{dwmptdomfin}
1 + \omega_{\rm m}^2 \tau^2 \lesssim |\beta|,
\end{equation}
and for the optomechanical damping when
\begin{equation}
\label{dGmptdomfin}
1 + \omega_{\rm m}^2 \tau^2 \lesssim \frac{\left| \beta \right| \kappa \tau}{2}.
\end{equation}
We note the difference of $\kappa \tau / 2$ between Eqs.~\eqref{dwmptdomfin} and \eqref{dGmptdomfin}, as this factor becomes significant when determining which optomechanical force dominates each of these dynamical back-action effects. This is especially true for non-sideband-resolved optomechanical cavities, where $\kappa \tau$ tends to be large and photothermal damping effects are generally stronger than those found in sideband-resolved systems \cite{pinard_2008, restrepo_2011, abdi_2012a, abdi_2012b}.

In Fig.~\ref{fig1}, we investigate three different optomechanical regimes according to Eqs.~\eqref{dwmptdomfin} and \eqref{dGmptdomfin}. Interestingly, we find that there exists a parameter space where $|\beta| < 1 + \omega_{\rm m}^2 \tau^2$, such that the spring effect is dominated by the radiation pressure force, but $\kappa \tau$ is large enough such that Eq.~\eqref{dGmptdomfin} is satisfied and optomechanical damping is governed by the photothermal force. This is particularly interesting in the case where $\beta$ is negative, resulting in the bizarre effect of a qualitatively similar detuning dependencies between the optomechanical damping and spring effect, as seen in Fig.~\ref{fig1} (c)/(f). It is this regime, which we refer to as the ``dueling regime,'' that we investigate experimentally in the remainder of the paper.

\section{Experiment}
\label{experiment}

\begin{figure}[t!]
\centerline{\includegraphics[width=\columnwidth]{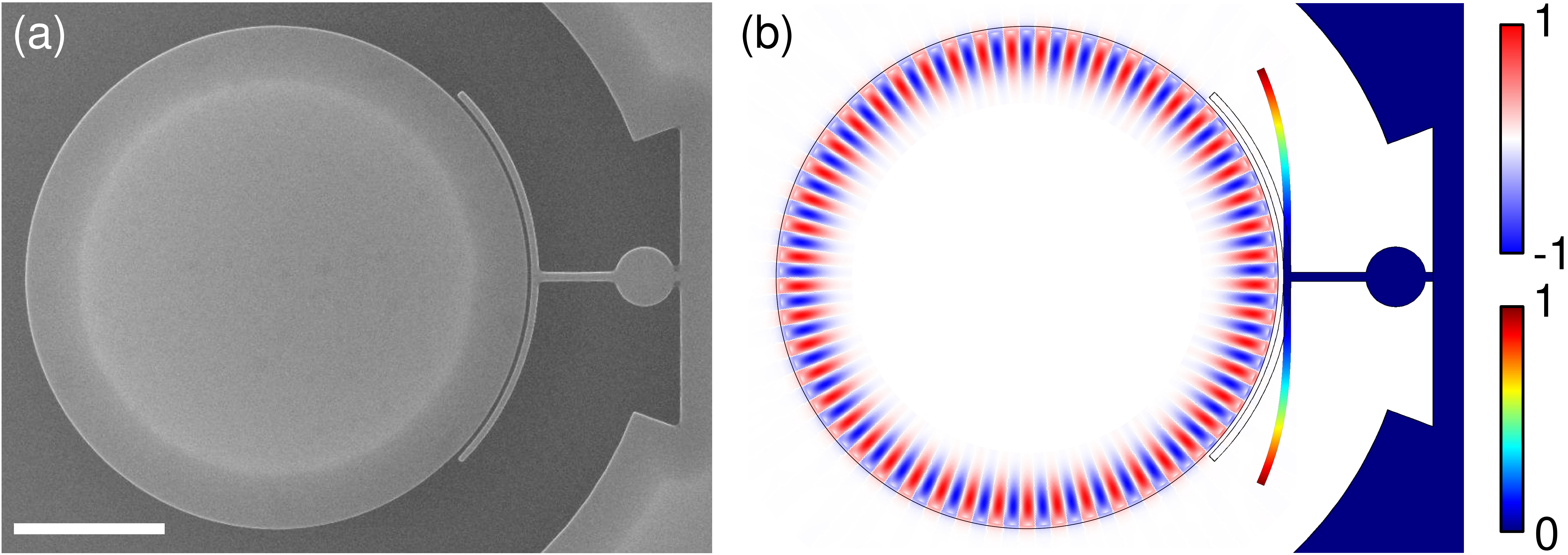}}
\caption{{\label{fig2}} (a) A scanning electron microscope image of the device studied in this work. Scale bar is 3 $\mu$m (see Appendix \ref{expdet} for detailed device dimensions). (b) Finite element method simulations of the normalized electric field magnitude for the first-order radial optical whispering-gallery mode of the disk with azimuthal mode number $M$ = 49. Colors indicate the direction of the in-plane electric field, with blue/negative (red/positive) corresponding to an inward (outward) facing field with respect to the center of the disk. Also included is a finite element method simulation of the in-plane flexural crab mode of the mechanical resonator with the normalized displacement expressed in rainbow scale. Both mechanical and optical simulations are for the device in (a).}
\end{figure}

The optomechanical device studied in this work is comprised of a ``claw-like'' mechanical resonator that surrounds one quarter of the circumference of a 10 $\mu$m diameter microdisk (see Fig.~\ref{fig2}). Both elements are fabricated from the 250 nm thick single-crystal silicon device layer of a standard SOI chip (fabrication details can be found elsewhere \cite{kim_2016}). The microdisk supports whispering-gallery-mode resonances in the telecom band, while the mechanical element exhibits a number of MHz-frequency flexural and torsional modes [see Fig.~\ref{fig2}(b) for example]. For this device geometry, dispersive coupling arises between the optical and mechanical modes of the system due to the fact that the resonator's motion through the evanescent field surrounding the microdisk acts to modulate its effective index, and therefore its optical resonance frequencies. In this work, we focus on the in-plane flexural ``crab'' mode of the mechanical resonator [see Fig.~\ref{fig2}(b)] with a measured resonant frequency of $\omega_{\rm m} / 2 \pi$ = 11.2 MHz, as this mode traverses the steepest gradient of the optical field profile, resulting in a large optomechanical coupling of $G/2 \pi$ = 0.82 GHz/nm. Using the measured dimensions of the mechanical resonator (see Appendix \ref{expdet}), along with its simulated modeshape, we find the effective motional mass of this mode to be $m$ = 183 fg \cite{hauer_2013}, allowing us to determine the zero-point fluctuation amplitude of its ground state as $x_{\rm zpf} = \sqrt{\hbar / 2 m \omega_{\rm m}}$ = 64 fm.

All measurements of the device are performed inside a cryostat using a custom-built cryogenic optomechanical coupling apparatus \cite{macdonald_2015}, with exchange gas added to the vacuum can to ensure thermalization of the device to the helium bath temperature of 4.2 K. Using this setup, laser light is directly injected into, and collected from, the optical cavity using a cryogenic dimpled-tapered fiber \cite{michael_2007, hauer_2014}. The DC transmission through the optical cavity is then monitored by directly observing the laser fluence through the fiber using a photodetector, while AC fluctuations in the optical signal are either transduced directly using this photodetector, or by switching out to a homodyne detection system (see Appendix \ref{expdet} for details). This allows for the advantage of being able to measure the mechanical signal using both direct and homodyne detection, as these two schemes are complimentary in a sense that one's response will be maximized for detunings at which the other is minimized, allowing for optimal signal-to-noise in the transduced signal over the entire sweep of the optical resonance.

\begin{figure}[t!]
\centerline{\includegraphics[width=\columnwidth]{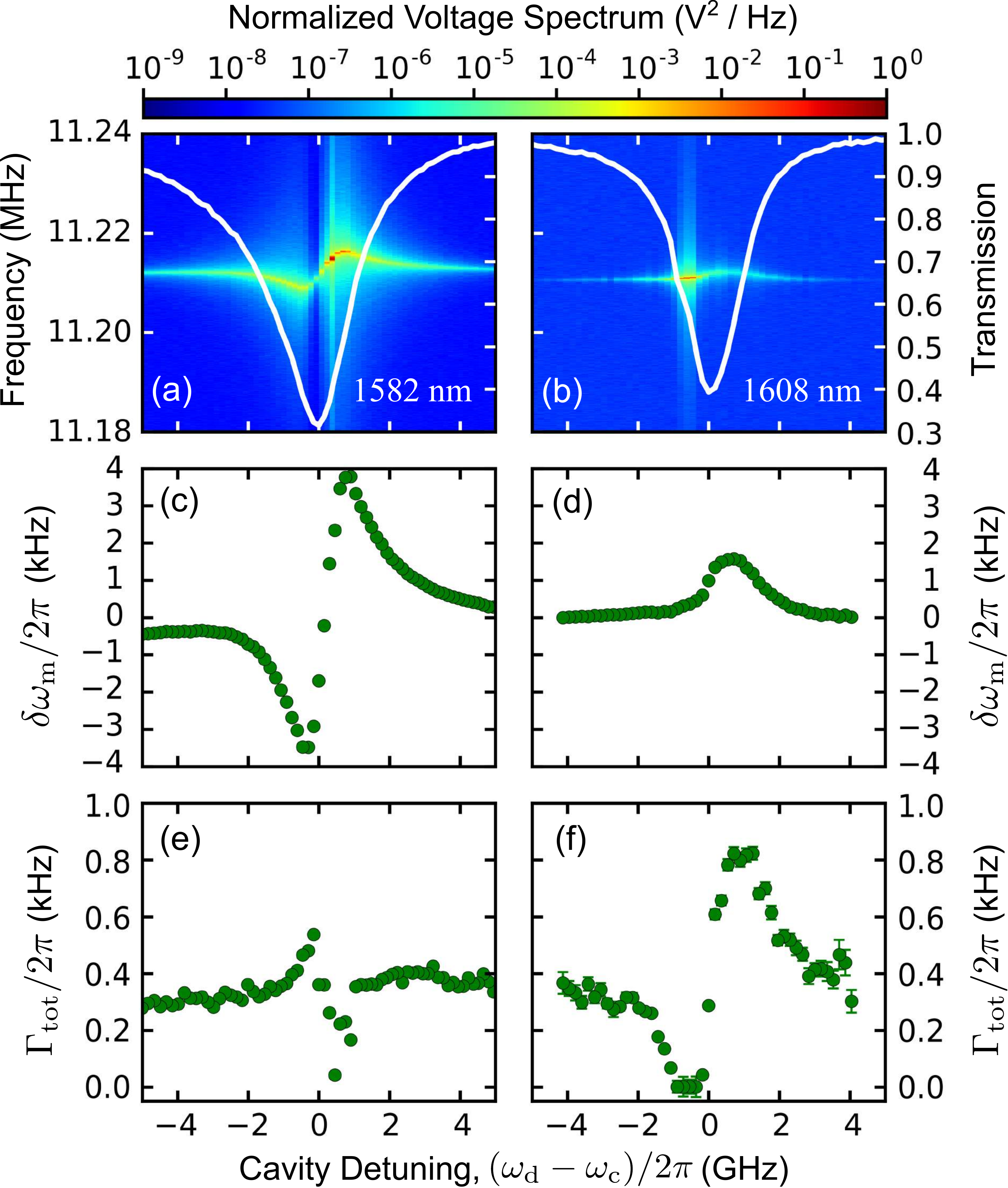}}
\caption{{\label{fig3}} Mechanical spectra (normalized to their maximum values) versus detuning for the 11.2 MHz crab mode depicted in Fig.~\ref{fig2}(b) for optical resonances at (a) $\lambda_{\rm c}$ = 1582 nm and (b) $\lambda_{\rm c}$ = 1608 nm, with the mechanical frequency on the left axis. Overlaid in white is the DC transmission (right axis) for each optical mode. Note that while in (a) mechanical amplification (damping) occurs on the red (blue) side of the optical resonance, this effect is reversed in (b). Also included are (c)/(d) the optomechanical spring effect $\delta \omega_{\rm m}$ and (e)/(f) the total mechanical damping rate $\Gamma_{\rm tot} = \Gamma_{\rm m} + \delta \Gamma_{\rm m}$, with (c) and (e) corresponding to the mechanical data in (a), while (d) and (f) are extracted from the data in (b). We attribute the lack of spring effect on the red side of the optical resonance in (d) to an optical-heating-induced mechanical frequency shift that offsets the dynamical backaction effects. Measurements are taken at input optical powers to the cavity of (a)/(c)/(e) $P_{\rm in}$ = 10.9 $\mu$W and (b)/(c)/(f) $P_{\rm in}$ = 1.9 $\mu$W, chosen such that self-oscillation of the mechanical motion has just begun to onset for each optical mode.}
\end{figure}

\section{Results}
\label{results}

In Fig.~\ref{fig3}, we show measurements of the studied optomechanical system from two separate optical resonances with center wavelengths located at 1582 nm [Fig.~\ref{fig3}(a)] and 1608 nm [Fig.~\ref{fig3}(b)]. Here the optical resonance at 1582 nm exhibits the behavior one would expect for a standard radiation-pressure-driven optomechanical system [see Fig.~\ref{fig1}(a)/(d)], where we observe optomechanical damping on the red side of the optical cavity and amplification on the blue side, with the mechanical spring effect exhibiting the opposite detuning dependence. However, this is not the case for the optical resonance at 1608 nm. Instead, the optomechanical damping behaves quite differently, with amplification on the red side of the optical cavity and damping on the blue side. Furthermore, the spring effect seems to qualitatively follow the same detuning dependence as the optomechanical damping, such that these two dynamical backaction effects appear to violate the Kramers-Kronig relations \cite{aspelmeyer_2014b}. We note that this reversal in the detuning dependence of the optomechanical damping is observed at optical input powers down to 10 nW (see Appendix \ref{powdep}), indicating that this seemingly anomalous effect does not onset at a given power threshold. We attribute this behavior to an additional photothermal force that is present for the 1608 nm optical mode, with $\beta$ and $\tau$ satisfying $\beta < 0$ and Eq.~\eqref{dGmptdomfin}, but not Eq.~\eqref{dwmptdomfin}, such that this force acts to overwhelm the device's radiation-pressure-driven optomechanical damping, but not its spring effect. We postulate that photothermal effects arise in this optomechanical device for optical modes that heat the inner surface of the mechanical resonator (facing the disk) via optical absorption. This in turn generates a thermal gradient across the width of the claw portion of the resonator, causing it to curl due to thermoelastic forces, thus actuating the crab mode.

\begin{figure}[t!]
\centerline{\includegraphics[width=\columnwidth]{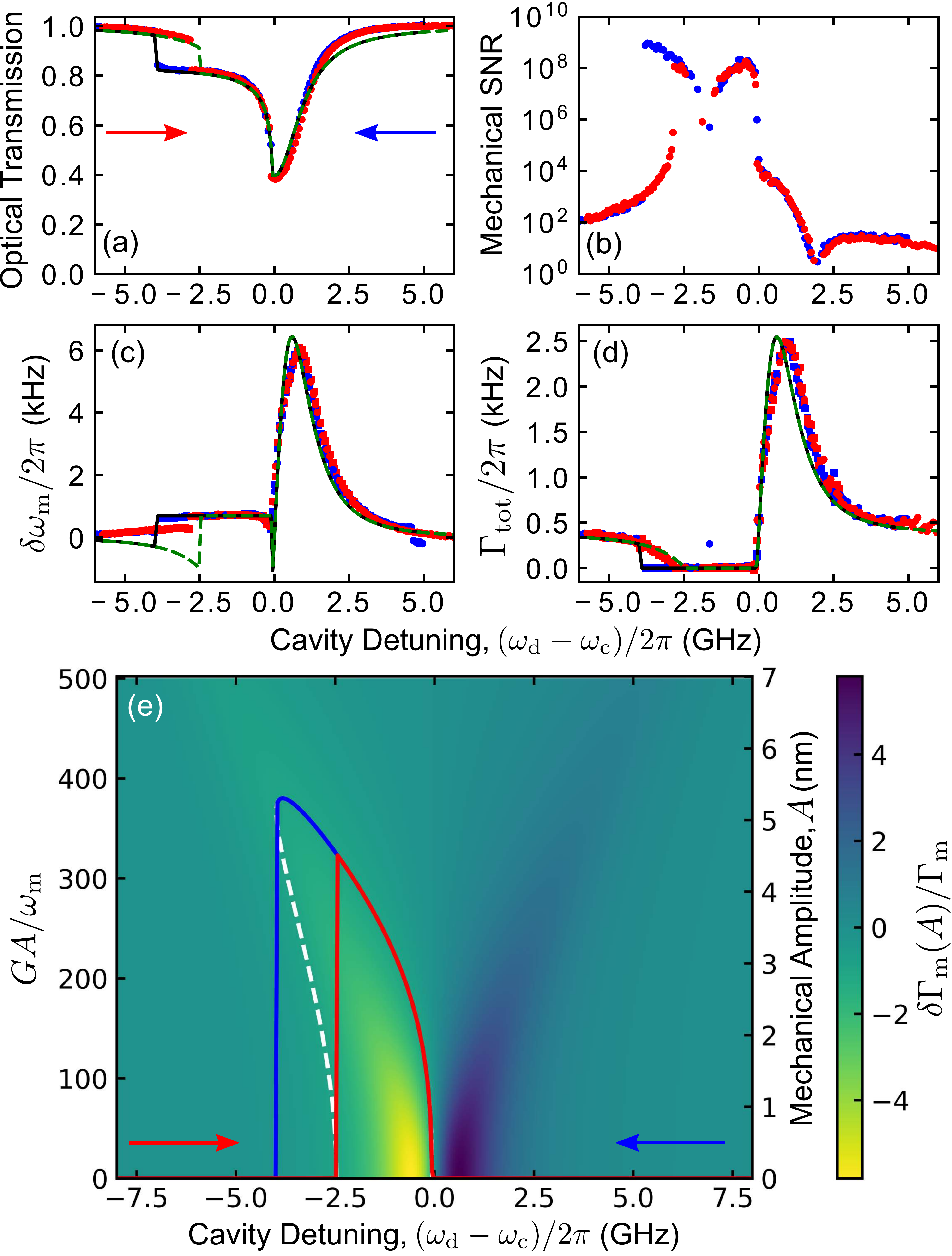}}
\caption{{\label{fig4}} At an input power of $P_{\rm in}$ = 10.1 $\mu$W, we find that the photothermal force drives the mechanical resonator into self-oscillation, causing an increase in its amplitude of motion. This results in a highly nonlinear response for (a) the DC transmission through the optical cavity and (b) the signal-to-noise ratio (SNR) of the homodyne mechanical spectra, as well as the optomechanical (c) spring effect and (d) damping over the detuning range from $\Delta \sim$ 0 to -4 GHz. In each of these plots, red (blue) data points denote an optical drive that was swept starting from the red (blue) side of the optical cavity, {\it i.e.}, from negative to positive (positive to negative) detunings, as indicated by the arrows in (a) and (e). The data in (c) and (d) are extracted from mechanical spectra obtained using both direct (circles) and homodyne (squares) detection of the high-frequency portion of the optical signal at each cavity detuning, while the signal-to-noise ratio in (b) is determined by dividing the maximum value of the homodyne spectra by its off-resonant imprecision noise floor. The data required for each detuning takes approximately 5 s to acquire, such that the sweep over the entire $\sim$160 detunings occurs on the timescale of $\sim$800 s. The dashed green (solid black) lines in (a), (c) and (d) are fits to the red (blue) data using Eqs.~\eqref{Tdcnl}, \eqref{dwmnl} and \eqref{dGmnl}, respectively, allowing for extraction of the optomechanical parameters quoted in the main text. In (e), we display an attractor diagram produced using Eq.~\eqref{dGmnl} to calculate $\delta \Gamma_{\rm m}(A)$ for a number of mechanical amplitudes and cavity detunings. The white dashed line indicates the condition of $\delta \Gamma_{\rm m}(A) / \Gamma_{\rm m} = -1$ ({\it i.e.}, $\Gamma_{\rm tot} = 0$), such that the red (blue) solid line traces out the physical values of the mechanical amplitude for a detuning sweep originating on the red (blue) side of the optical cavity.}
\end{figure}

The inverted detuning dependence associated with this effect becomes more pronounced at higher optical input powers, where we further find that the observed photothermal amplification is strong enough to reduce the total mechanical damping of the system to zero, inducing a parametric instability \cite{marquardt_2006, ludwig_2008}. This causes the device to self-oscillate for a near-resonant red-detuned optical pump, driving the mechanical resonator's motion to amplitudes as large as $A_{\rm max}$ = 5.3 nm [= 380 $\omega_{\rm m} / G$ -- see Fig.~\ref{fig4}(e)], nearly three orders of magnitude greater than its thermally-driven amplitude of $A_{\rm th} = \sqrt{2 k_{\rm B} T / m \omega_{\rm m}^2} =$ 11.6 pm at $T$ = 4.2 K.  Accompanying this increase in mechanical amplitude, we also observe highly nonlinear behavior in each of the spring effect, optomechanical damping, and DC transmission through the optical cavity (see Fig.~\ref{fig4}), as well as a hysteresis in each of these quantities with respect to the sweep direction of the optical drive. We note that while the optomechanical interaction causing the device to enter self-oscillations is nonlinear, the mechanical motion itself is still remains within the linear regime, avoiding complications such as Duffing nonlinearities \cite{lifshitz_2008}.

This peculiar behavior can be understood by examining the combined photothermal and radiation pressure attractor diagram of the system \cite{marquardt_2006, ludwig_2008, metzger_2008b, krause_2015, buters_2015}, which is generated by evaluating Eq.~\eqref{dGmnl} at various mechanical amplitudes $A$ and optical cavity detunings $\Delta = \Delta_0 + G \bar{x}$ [see Fig.~\ref{fig4}(e)]. Note that with this definition of $\Delta$, we include the shift in the cavity resonance due to the static, optomechanically induced displacement of the mechanical resonator. In principle, this static shift can act to displace the detuning dependence of the attractor diagram \cite{metzger_2008b}, however, for the device considered here this shift is negligible, such that $\Delta \approx \Delta_0$ (see Appendix \ref{omeoms}). The physical values for the mechanical amplitude are found to traverse the contours of the attractor diagram that obey $\Gamma_{\rm tot} = 0 \Rightarrow \delta \Gamma_{\rm m} = - \Gamma_{\rm m}$ [see white dashed line in Fig.~\ref{fig4}(e)], corresponding to an increase in mechanical amplitude in order to dissipate the optical power input to the system \cite{marquardt_2006, ludwig_2008}. As can be seen in Fig.~\ref{fig4}(e), for the non-sideband-resolved system considered here, there are two possible mechanical amplitude solutions for cavity detunings ranging from $\Delta \approx$ -2.5 to -4.0 GHz, leading to dynamical bistability and therefore a hysteresis in the mechanical amplitude, as well as the optomechanical properties of the system \cite{metzger_2008b}. We point out that these nonlinear effects, which are a result of the photothermal force, would not be present for this system if only the radiation pressure force were considered (see Appendix \ref{radatt}).

Fixing the mechanical resonance frequency and damping rate to their low power values of $\omega_{\rm m} / 2 \pi$ = 11.2 MHz and $\Gamma_{\rm m}/ 2 \pi$ = 374 Hz, while using $\tau$ = 9.1 ns determined from finite-element method simulations (see Appendix \ref{tauthsec}), we fit the data in Fig.~\ref{fig4} by varying $G$, $\beta$, $\kappa$ and $\kappa_{\rm e}$. We note that while driven to self-oscillation, the mechanical frequency locks to a position slightly larger than its off-resonant value, which we attribute to a small thermal shift in the mechanical resonance due to optically induced heating of the resonator [see Fig.~\ref{fig4}(c)], leading to an additional, inconsequential fit parameter. To perform this fitting procedure, we first determine the mechanical amplitude according to an attractor diagram similar to that in Fig.~\ref{fig4}(e) for each iteration of guess parameters. These amplitude are then input into Eqs.~\eqref{dwmnl}, \eqref{dGmnl}, and \eqref{Tdcnl}, which are compared to the data in Figs.~\ref{fig4}(c), (d) and (a), respectively. This process is repeated until the minimization condition of the fitting algorithm is met (see Appendix \ref{omeoms} for more details). Using this procedure, we extract the optomechanical coupling parameters $G/2 \pi$ = 0.82 GHz/nm ($g_0 / 2 \pi = G x_{\rm zpf} / 2 \pi$ = 52.5 kHz) and $\beta$ = -0.318, along with the total optical loss rate of $\kappa/ 2 \pi$ = 2.04 GHz and the external coupling rate of $\kappa_{\rm e} / 2 \pi$ = 0.38 GHz, for the studied device. This results in a single photon cooperativity of $\mathcal{C}_0 = 4 g_0^2 / \kappa \Gamma_{\rm m}$ = 1.4 $\times$ 10$^{-2}$ and a maximal cavity-enhanced cooperativity of $\mathcal{C} = \bar{N}_{\rm max} \mathcal{C}_0$ = 68, where $\bar{N}_{\rm max} = 4 \kappa_{\rm e} P_{\rm in} / \hbar \omega_{\rm d} \kappa^2$ = 4.7 $\times$ 10$^3$ is the average number of photons circulating within the cavity for a resonant pump with an input power of $P_{\rm in}$ = 10.1 $\mu$W.  From these extracted parameters, we also find that $\omega_{\rm m} \tau$ = 0.64, leading to $1 + \omega_{\rm m}^2 \tau^2$ = 1.41 and $|\beta| \kappa \tau$ = 37. This ensures that such that Eq.~\eqref{dGmptdomfin} is satisfied, while Eq.~\eqref{dwmptdomfin} is not, confirming that we are indeed in the dueling regime associated with a radiation-pressure-dominated spring effect, but a photothermal-dominated optomechanical damping.

\begin{figure}[t!]
\centerline{\includegraphics[width=\columnwidth]{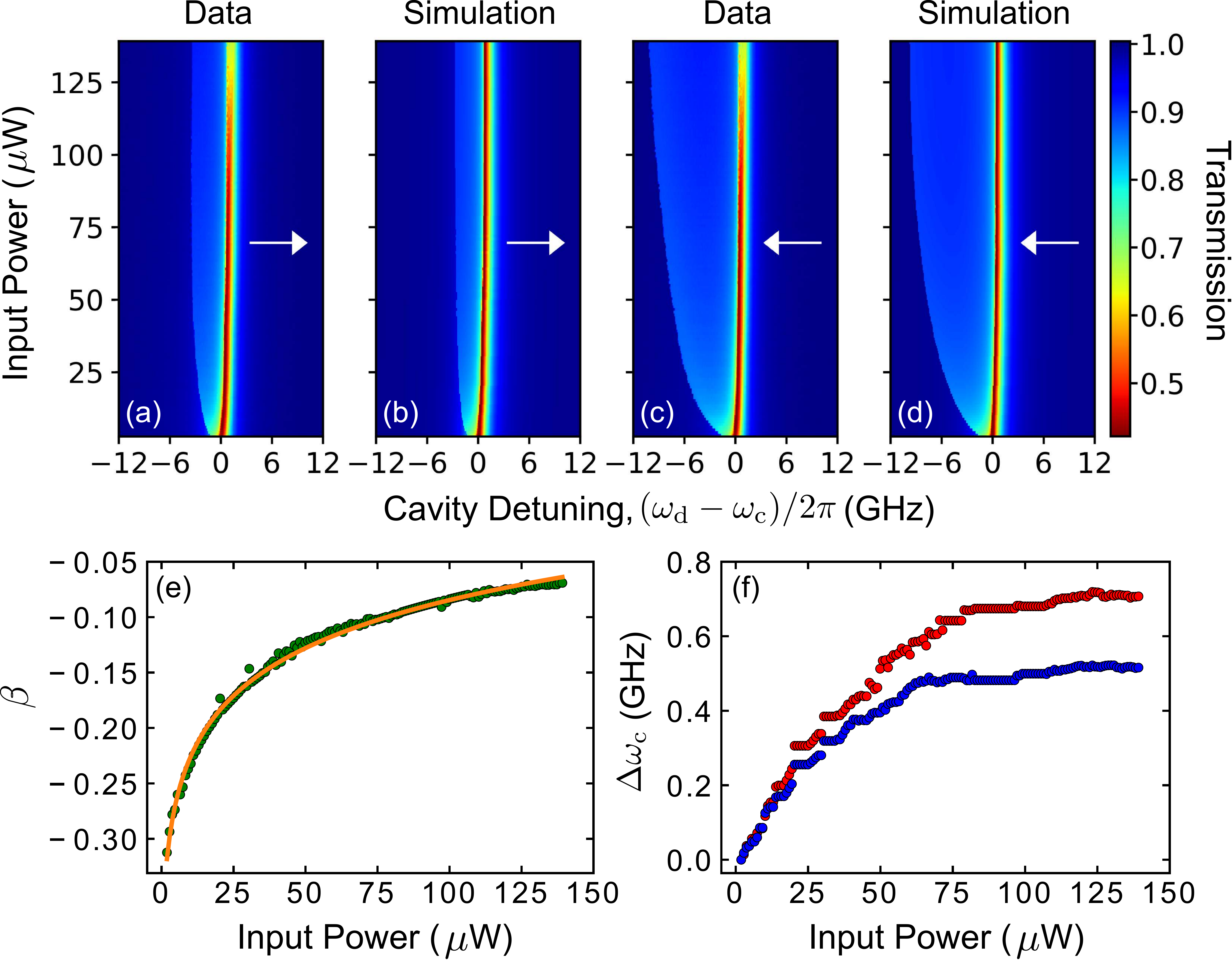}}
\caption{{\label{fig5}} (a)-(d) DC transmission through the optical cavity as a function of detuning and input power. Here we show the (a)/(c) data and (b)/(d) simulation for red and blue detuning sweeps, as indicated by the white arrow, with each detuning referenced to the cavity's resonance frequency at the lowest measured power of $P_{\rm in}$ = 1.9 $\mu$W. Fixing $\kappa / 2 \pi$ = 1.59 GHz and $\kappa_{\rm e} / 2 \pi$ = 0.29 GHz, we fit the data in (a)/(c) to Eq.~\eqref{Tdcnl}, allowing us to determine $\beta$ and $\omega_{\rm c}$ versus input optical power to the cavity. In (e), we display these values for $\beta$ (green circles), which are fit to Eq.~\eqref{BvP} (orange line) resulting in $\beta_0 = -0.409$, $B = 0.063$, and $P_*$ = 0.59 $\mu$W. The simulations in (b)/(d) are calculated using Eq.~\eqref{Tdcnl} with the values of $\beta$ from the fit in (e), along with $\omega_{\rm c}$ extracted from the fits to (a)/(c) and the aforementioned fixed values of $\kappa$ and $\kappa_{\rm e}$. Finally, (f) displays the optical cavity resonance frequency shift $\Delta \omega_{\rm c}$ (relative to its value at the lowest optical power), with the red/blue data points corresponding to fits of the detuning sweeps originating from the red/blue side of the cavity found in (a)/(c), both of which indicate a blue shift in the cavity resonance with increasing optical power.}
\end{figure}

We continue to observe nonlinear effects in the optical transmission through the cavity for input powers up to $\sim$140 $\mu$W, the data for which can be seen in Fig.~\ref{fig5}. These measurements were performed under coupling conditions that differed slightly from those used to collect the data for Fig.~\ref{fig4}, causing a shift in the loss rates of the optical cavity to $\kappa / 2 \pi$ = 1.59 GHz and $\kappa_{\rm e} / 2 \pi$ = 0.29 GHz. Fixing these values for $\kappa$ and $\kappa_{\rm e}$, as well as assuming that the thermal relaxation time remains constant in power/temperature (which should be the case up to roughly 100 K -- see Appendix \ref{tauthsec}), we fit the optical scans in Figs.~\ref{fig5}(a)/(c) to Eq.~\eqref{Tdcnl}, extracting $\beta$ and $\omega_{\rm c}$ versus input optical power to the cavity [see Figs.~\ref{fig5}(e)/(f)]. Upon inspection of Fig.~\ref{fig5}(e), we find that $\beta$ exhibits a logarithmic dependence on input power, which can be fit with the phenomenological equation
\begin{equation}
\label{BvP}
\beta(P) = \beta_0 + B \ln \left( 1 + \frac{P}{P_*} \right),
\end{equation}
where $\beta_0 = -0.409$ is the value of $\beta$ at zero power, while $B  = 0.063$ and $P_*$ = 0.59 $\mu$W are scaling parameters. Rearranging Eq.~\eqref{BvP}, we can also determine the power at which $\beta = 0$ as $P_0 =  P_* (e^{-\beta_0 / B} - 1)$ = 384 $\mu$W. Inputting $\beta$ from this fit, along with the extracted values of $\omega_{\rm c}$ versus power, into Eq.~\eqref{Tdcnl}, we show that we are able to reproduce the power-dependent behavior of the optical transmission data, as can be seen in Figs.~\ref{fig5}(b)/(d).

The observed power-dependence in $\beta$ and $\omega_{\rm c}$ is likely due to the fact that increasing the power input to the optomechanical system causes it to heat up, changing its thermal and optical properties. Due to the complicated nature of optically induced heating, it is difficult to quantitatively ascertain the temperature of the device in this regime, however, we find the qualitative trend that $\beta$ decreases in magnitude as we move to higher power/temperature. Furthermore, we observe that $\omega_{\rm c}$ increases with power/temperature, which is consistent with the negative thermal expansion coefficient observed for silicon between approximately 17 K and 120 K \cite{lyon_1977}, as an increase in temperature reduces the diameter of the microdisk cavity, resulting in a blue-shift in its optical resonant frequency \cite{carmon_2004, macdonald_2016}. We postulate that it is this decrease in the magnitude of $\beta$ with increasing temperature that has prevented previous studies of the dueling radiation pressure and photothermal effects discussed in this paper, as the majority of optomechanical experiments on nanophotonic silicon devices have been performed at room temperature. 

We conclude this section by noting that $\beta = -0.232$ at $P_{\rm in} =$ 10.2 $\mu$W for the data in Fig.~\ref{fig5}, which is considerably smaller in magnitude than the value of $\beta = -0.318$ extracted from Fig.~\ref{fig4}, where $P_{\rm in} =$ 10.1 $\mu$W. We attribute this disparity to the fact that while the power input to the optical cavity is similar in each case, due to the discrepancies in the optical linewidths between the two measurements, the power absorbed by the mechanical resonator, and therefore its temperature, is larger for the data in Fig.~\ref{fig5}, resulting in a decrease in the magnitude of $\beta$ (for a more indepth discussion of this effect see Appendix \ref{omeoms}).

\section{Optomechanical Cooling}
\label{cooling}

Up to this point we have largely focussed on the photothermally driven amplification of mechanical motion that occurs for a pump beam detuned to the red side of the optical resonance. However, this photothermal effect can also be used to perform considerable cooling of the mechanical mode on the opposite (blue) side of the resonance. For instance, in Fig.~\ref{fig4}(d) we find that the photothermal force increases the total damping rate of the mechanical resonator to as high as $\Gamma_{\rm tot} / 2 \pi$ = 2.5 kHz at $\Delta$ = $2 \pi$ $\times$ 0.98 GHz (= 0.48 $\kappa$), resulting in a factor of 6.7 increase from its intrinsic value of $\Gamma_{\rm m} / 2 \pi$ = 374 Hz. Assuming that the resonator is initially thermalized to the helium bath temperature of $T_{\rm b}$ = 4.2 K, this damping effect actively cools the mechanical mode to a temperature of $T_{\rm m} = T_{\rm b} (\Gamma_{\rm m} / \Gamma_{\rm tot})$ = 631 mK, equivalent to a reduction in the phonon occupation of the mechanical resonator from $\braket{n} \approx$ 7800 to $\braket{n} \approx$ 1170 \cite{aspelmeyer_2014}. This cooling effect is especially intriguing considering that it occurs for a blue-detuned optical pump, such that the photothermal force must overwhelm any radiation-pressure-driven amplification effects.

\begin{figure}[t!]
\centerline{\includegraphics[width=3.0in]{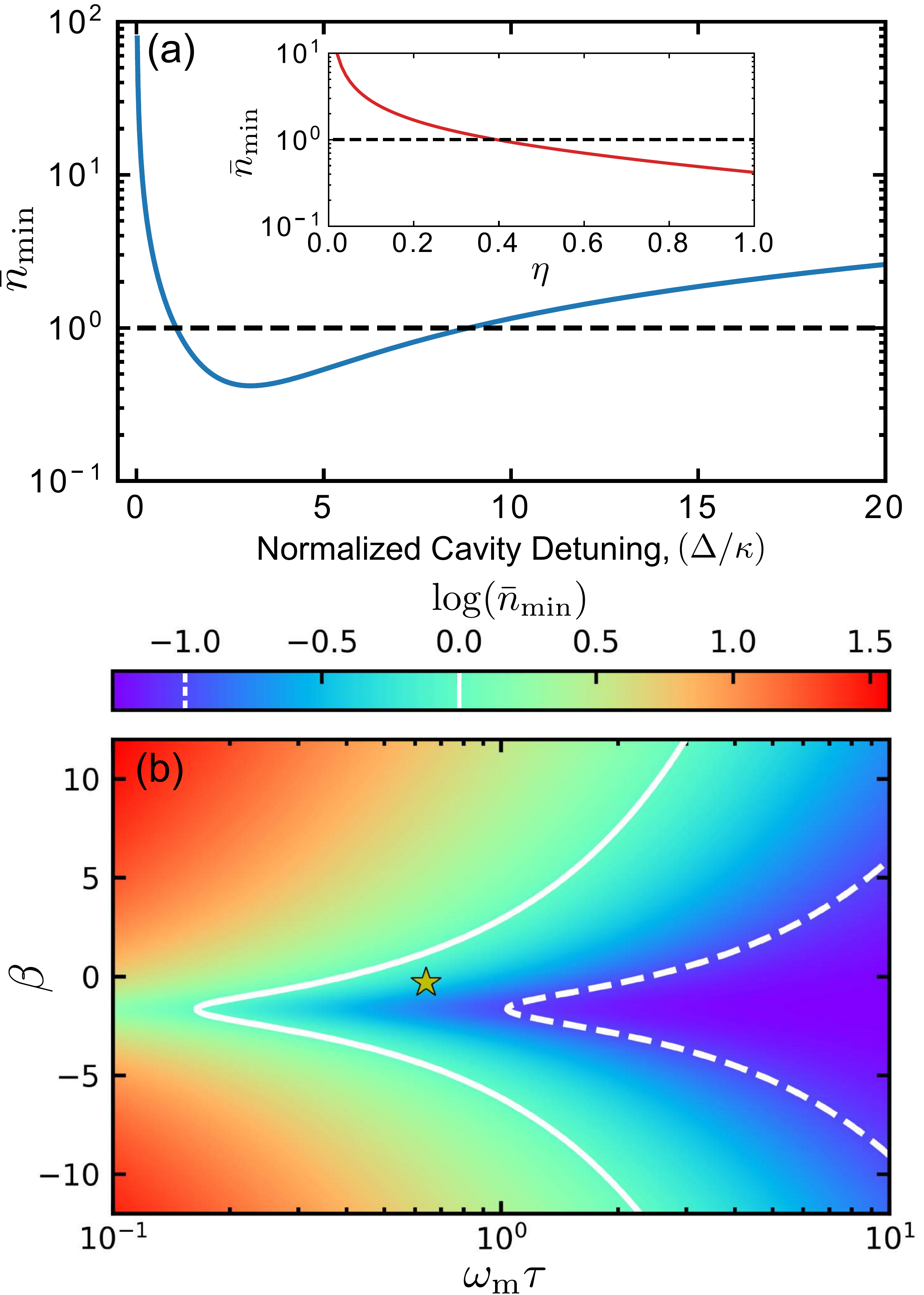}}
\caption{{\label{fig6}} (a) Plot of the minimum reachable phonon number according to Eq.~\eqref{nmin} versus detuning using the parameters extracted from the fits in Fig.~\ref{fig4}, while setting $\kappa_{\rm a} = \kappa_{\rm i}$. The black dashed line corresponds to $\bar{n}_{\rm min} = 1$, indicating that the resonator can in principle be cooled below single phonon occupancy over a detuning band from $\Delta \sim \kappa$ to $9 \kappa$, reaching its minimum value of $\bar{n}_{\rm min}$ = 0.42 at $\Delta_{\rm min}$ = 3.0 $\kappa$. Inset is a plot of $\bar{n}_{\rm min}$ as a function of the ratio $\eta = \kappa_{\rm a} / \kappa_{\rm i}$, indicating that ground state cooling is possible for $\eta \gtrsim 0.4$. (b) A color plot depicting the base-ten logarithm of $\bar{n}_{\rm min}$, minimized with respect to detuning, as a function of $\beta$ and $\omega_{\rm m} \tau$, with the rest of the parameters the same as in (a). Here the solid (dashed) white lines demarcate the contour of $\bar{n}_{\rm min} = 1$ ($\bar{n}_{\rm min} = 0.1$), while the star indicates the parameters for the device studied here. As one can see, there is a region where $\bar{n}_{\rm min} < 1$ centered near $\beta = 0$, with deviations from a symmetric distribution in $\beta$ being due to interference between the radiation pressure and photothermal forces \cite{restrepo_2011}. We further note that while $\bar{n}_{\rm min}$ decreases for larger values of $\omega_{\rm m} \tau$, the detuning for which $\bar{n}_{\rm min}$ is minimized increases with $\omega_{\rm m} \tau$ (see Appendix \ref{omeoms}), moving away from $\Delta \approx \kappa /2$ where the photothermal force is maximal, such that it becomes increasingly difficult to reach $\bar{n}_{\rm min}$ experimentally.}
\end{figure}

More interesting, however, is the fundamental limit on minimum reachable phonon number using this cooling mechanism, which is set by the shot noise generated by photons impinging upon the mechanical resonator. For a purely radiation-pressure-driven system, this limit is given by
\begin{equation}
\label{nminrp}
\bar{n}_{\rm min}^{\rm rp} = - \frac{\frac{\kappa^2}{4} + (\Delta + \omega_{\rm m})^2 }{ 4 \Delta \omega_{\rm m}}, 
\end{equation}
which when minimized with respect to detuning in the non-sideband-resolved regime ($\kappa \gg \omega_{\rm m}$) results in $\bar{n}_{\rm min}^{\rm rp} \approx \kappa / 4 \omega_{\rm m}$ \cite{marquardt_2007, aspelmeyer_2014}. It is important to note that this result only holds true for standard, dispersively coupled optomechanical cavities, as it has been shown that ground state cooling can in theory be achieved using non-sideband-resolved, dissipatively coupled systems \cite{elste_2009}. Nonetheless, we find that $\bar{n}_{\rm min}^{\rm rp} \approx$ 45 for the device studied here, such that it would be impossible to cool it to an average phonon occupation less than one using radiation pressure alone. However, the situation is far more complex when one adds photothermal effects into the picture, as this force interferes with the radiation pressure \cite{restrepo_2011}, resulting in a modified expression for the minimum achievable phonon number given by (see Appendix \ref{omeoms})
\begin{equation}
\begin{split}
\label{nmin}
&\bar{n}_{\rm min} = -\frac{ \frac{\kappa^2}{4} + (\Delta + \omega_{\rm m})^2 }{4 \Delta \omega_{\rm m} \left\{\kappa + \frac{\beta}{1 + \omega_{\rm m}^2 \tau^2} \left[ \kappa + \tau \left( \frac{\kappa^2}{4} + \Delta^2 -\omega_{\rm m}^2 \right) \right] \right\}} \\
&\times \bigg\{ \kappa +  \frac{\beta}{1 + \omega_{\rm m}^2 \tau^2} \bigg[ \kappa \left( \frac{\beta \kappa}{4 \kappa_{\rm a}} + 1 \right) \\
&+ \left(\Delta + \omega_{\rm m} \right) \left( \frac{\beta (\Delta + \omega_{\rm m})}{\kappa_{\rm a}} - 2 \omega_{\rm m} \tau \right) \bigg] \bigg\}.
\end{split}
\end{equation}
Here we have introduced $\kappa_{\rm a} = \eta \kappa_{\rm i}$ as the optical loss rate due to absorption of photons in the mechanical resonator, which makes up a fraction $\eta$ of the total intrinsic loss rate of the cavity $\kappa_{\rm i}$. For the experimental measurements given in Fig.~\ref{fig4}, we determine this total intrinsic loss rate to be $\kappa_{\rm i} = \kappa - \kappa_{\rm e}$ = 1.66 GHz. It is difficult to experimentally determine what fraction of this intrinsic loss rate contributes to $\kappa_{\rm a}$, however, we initially assume that the optical loss rate is dominated by absorption in the mechanical element ({\it i.e.}, set $\eta = 1$), allowing us to set a lower limit on $\bar{n}_{\rm min}$ for the device studied here. Using this condition, along with the extracted experimental values from Fig.~\ref{fig4}, we plot the minimum achievable phonon number given by Eq.~\eqref{nmin} as a function of detuning in Fig.~\ref{fig6}(a). As one can see, the minimum achievable phonon number drops below one over a detuning range from $\Delta \sim \kappa$ to $9 \kappa$, reaching its minimum value of $\bar{n}_{\rm min}$ = 0.42 at $\Delta_{\rm min}$ = 3.0 $\kappa$, corresponding to a mechanical resonator that is in its ground state 70\% of the time. We note that ground state cooling remains possible when relaxing the condition that $\kappa_{\rm a} = \kappa_{\rm i}$, with $\bar{n}_{\rm min} < 1$ for $\eta \gtrsim 0.4$ [see inset of Fig.~\ref{fig6}(a)]. While it has long been known theoretically that the photothermal force can be used to cool a non-sideband-resolved optomechanical resonator into its motional ground state \cite{pinard_2008, deliberato_2011, restrepo_2011, abdi_2012a, abdi_2012b}, this is the first time that a device has been experimentally demonstrated to exist within the required regime.

To further investigate the parameter space over which ground state photothermal cooling can occur, we have plotted the logarithm of the minimum achievable phonon number versus $\beta$ and $\omega_{\rm m} \tau$ in Fig.~\ref{fig6}(b). Each point on this plot is obtained by varying $\beta$ and $\tau$ in Eq.~\eqref{nmin} (while again setting all other physical parameters equal to those extracted from the fits to Fig.~\ref{fig4}) and taking the minimum value of $\bar{n}_{\rm min}$ with respect to detuning. The result is a large region of photothermal parameter space that allows for cooling below the single phonon level, with a slight asymmetry between positive and negative $\beta$ due to interference between the radiation pressure and photothermal forces \cite{restrepo_2011}. As indicated by the yellow star in Fig.~\ref{fig6}(b), the parameters for the device considered in this work lie well within this regime. 

One must be careful, however, when interpreting these results, as $\bar{n}_{\rm min}$ describes the fundamental limit on the minimum reachable phonon number using this cooling mechanism. Furthermore, as $\omega_{\rm m} \tau$ increases, so does the detuning at which $\bar{n}_{\rm min}$ is minimized, reducing the effectiveness of the photothermal cooling (see Appendix \ref{omeoms}). Therefore, one generally wishes to maximize the strength of the photothermal damping force, which occurs when $\omega_{\rm m} \tau \approx 1$ \cite{deliberato_2011, metzger_2004, metzger_2008a}, in order to decrease the optical power required to reach $\bar{n}_{\rm min}$. Of particular interest are the photothermal cooling parameters of $\omega_{\rm m} \tau = 1$ and $\beta = -2.0$, which for the other device parameters used in this work, results in $\bar{n}_{\rm min} = 0.11$ at $\Delta_{\rm min} \approx 0.5 \kappa$ (see Appendix \ref{omeoms}). These conditions therefore maximize photothermal cooling with respect to both thermal relaxation time and detuning, while still allowing for ground state cooling of the mechanical resonator, thus presenting a set of parameters to strive for in future iterations of the device.

\section{Conclusion}
\label{conclusion}

In this paper, we have presented measurements of a silicon whispering-gallery-mode optomechanical cavity that exhibits dynamical backaction effects due to competing photothermal and radiation pressure forces. We find that the radiation pressure force governs the optomechanical spring effect, while the photothermal force dictates the optomechanical damping. Furthermore, due to the fact that this photothermal force acts to directly oppose its radiation pressure counterpart, we find that at high enough power we can reduce the mechanical damping to zero on the red side of the cavity resonance, inducing a parametric instability in the mechanical resonator that drives its motion into large amplitude self-oscillation. At the onset of this self-oscillating behavior, we observe highly nonlinear effects, as well as a hysteresis depending on the sweep direction of the optical drive, in each of the optomechanical damping, spring effect and DC transmission through the optical cavity. Fitting these data with a nonlinear optomechanical model that includes both radiation pressure and photothermal forces, we extract the optomechanical properties of the system associated with each of these effects. Finally, using these extracted parameters, we infer that this non-sideband-resolved optomechanical system can in principle be cooled to an average phonon occupancy less than one. This comprehension of exactly how the radiation pressure and photothermal forces interact with each other at low temperatures will be crucial as silicon optomechanical cavities continue to be used to perform quantum experiments \cite{cohen_2015, riedinger_2016, hong_2017, riedinger_2018, marinkovic_2018}.

While the ability to cool below single phonon occupancy in the non-sideband-resolved regime is promising, reaching this limit in practice presents a significant challenge, largely due to residual heating from the photon absorption processes associated with photothermal damping \cite{deliberato_2011}. However, as this device was not purposefully designed for photothermal coupling, it may be possible to engineer this effect to achieve the photothermal parameters detailed at the end of the previous section, perhaps by adding a metallic layer to the resonator to enhance its differential thermal contractions and optical absorption \cite{dealba_2017}. Furthermore, one could also imagine modifying the thermal time constant by changing the dimensions of the resonator, which would also affect the strength of the photothermal damping. Increasing the photothermal coupling in this way may provide a path to cool a photothermal optomechanical device into its motional ground state, as well as allow for future investigation of other photothermally enhanced optomechanical effects, such as entanglement \cite{abdi_2012a, abdi_2012b} or induced chaos \cite{marino_2011, marino_2013} between the optical and mechanical modes of the system.

\begin{acknowledgments}

The authors would like to thank A.~Metelmann, A.~Clerk, C.~Simon, and P.~Barclay for valuable discussions. This work was supported by the University of Alberta, Faculty of Science; the Natural Sciences and Engineering Research Council, Canada (Grants Nos. RGPIN-04523-16, DAS-492947-16, and CREATE-495446-17); and the Canada Foundation for Innovation.  B.D.H. acknowledges support from the Killam Trusts. 

\end{acknowledgments}

\begin{appendix}

\section{Experimental Details}
\label{expdet}

\subsection{Experimental Setup}
\label{expset}

\begin{figure*}[t!]
\centerline{\includegraphics[width=6in]{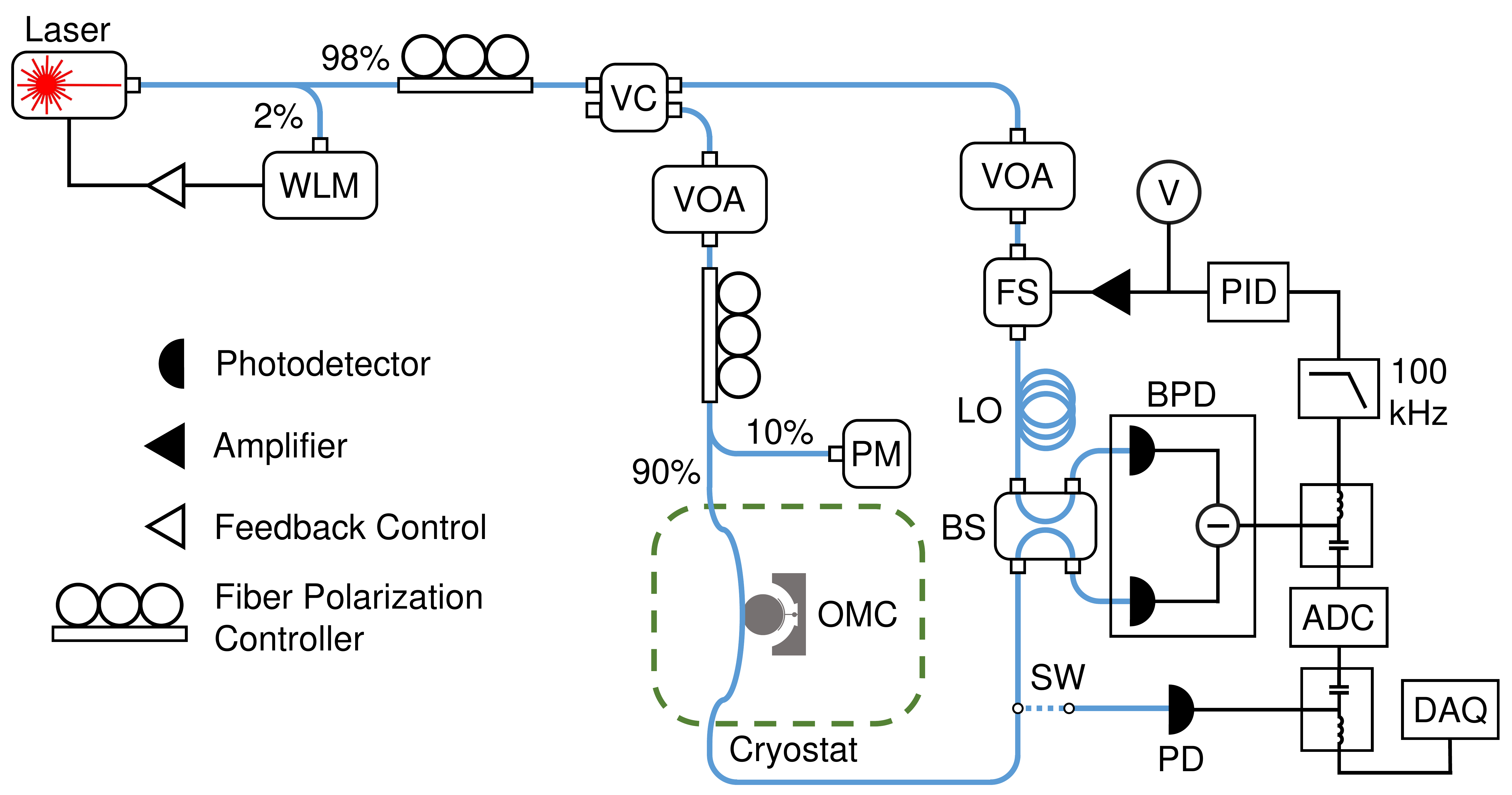}}
\caption{{\label{fig7}} Schematic of the optical detection system used to probe the optical and mechanical properties of the device under study. WLM = wavelength meter, VC = variable coupler, VOA = variable optical attenuator, PM = power meter, OMC = optomechanical cavity, FS = fiber stretcher, PID = proportional-integral-derivative controller, V = voltmeter, LO = local oscillator, SW = optical switch, BS = beam splitter, BPD = balanced photodetector, PD = photodetector, ADC = analog-to-digital converter, DAQ = data acquisition.}
\end{figure*}

In order to address the optomechanical device studied in this work, we use a cryogenic optical detection system (shown schematically in Fig.~\ref{fig7}) that allows for both direct detection and homodyne measurements of the collected optical signal. Light from a fiber-coupled tunable diode laser (1550 - 1630 nm), whose wavelength is stabilized on long timescales (on the order of hours) using a 2\% pickoff to a wavelength meter (WLM), is sent to a variable coupler (VC) that splits the optical circuit into two paths: the signal arm and the local oscillator (LO). The optical power in the signal arm is set using a voltage-controlled variable optical attenuator (VOA), all while being monitored by sending 10\% of this signal to a power meter (PM). This path continues through a fiber polarization controller, ensuring that the laser light headed to the optical cavity is polarization-matched to the optical mode of interest. After these components, the laser in the signal arm is then directed via optical fiber to a low temperature optomechanical coupling apparatus that resides on the base plate of a dilution refrigerator \cite{macdonald_2015}, complete with a dimpled optical fiber taper \cite{michael_2007, hauer_2014} that allows photons to couple to and from the on-chip optomechanical device. The intracavity signal is recollected using this tapered fiber and sent to an optical switch (SW) that toggles this signal between a standard photodetector (PD) for direct detection of the optical signal, or alternatively, to a balanced photodetector (BPD) for homodyne measurements. In the latter case, light from the signal arm is recombined with the LO on a fiber-coupled beam splitter (BS) and sent to the BPD, allowing for a phase-sensitive probe of the optical signal. The constant phase offset between the signal and LO arms is maintained using a proportional-integral-derivative (PID) controlled fiber stretcher (FS) located in the LO, with its setpoint referenced to the low frequency (DC) component of the difference signal from the BPD, which is monitored in real-time using a voltmeter (V). For each of the homodyne and direct detection setups, the high frequency (AC) signal is recorded as time-series data using a 500 MS/s analog-to-digital converter (ADC), allowing for observation of the mechanical motion. Finally, the DC transmission through the optical cavity is obtained by monitoring the low frequency ($<$25 kHz) channel of the direct detection PD using a data acquisition (DAQ) system.

\newpage

\subsection{Device Dimensions}
\label{devdim}

\begin{figure}[h!]
\centerline{\includegraphics[width=2.8in]{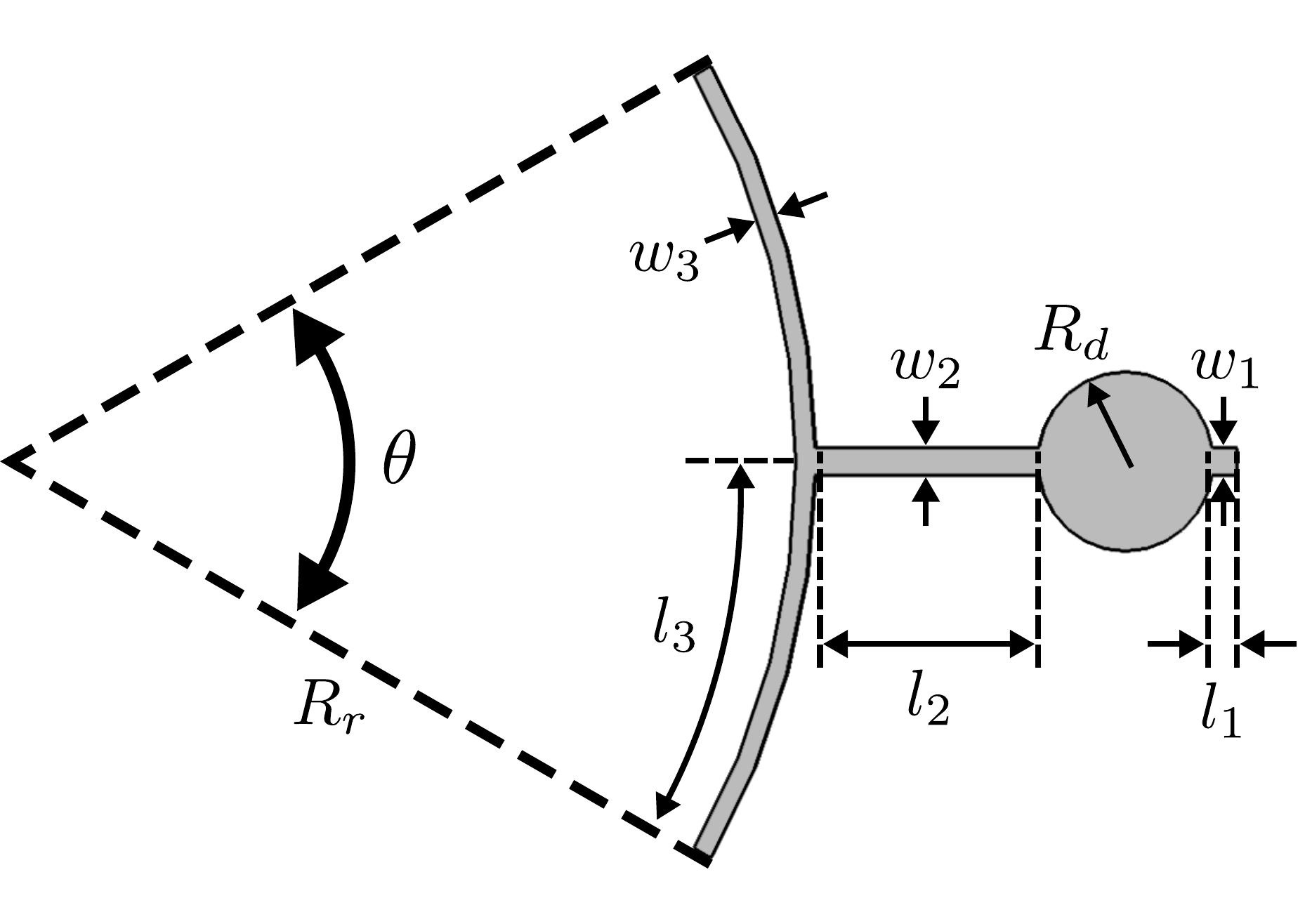}}
\caption{{\label{fig8}} Schematic of the studied optomechanical device, indicating each of its critical dimensions. Numerical values for each dimension are given in Table \ref{tab1}.}
\end{figure}

\begin{table}[h!]
\begin{tabular}{ ccc }
\hline
\multicolumn{3}{c}{Measured Dimensions} \\  
\hline
\hline
$l_1$ = 143 nm~~~~~~ & $w_1$ = 177 nm~~~~~~ & $R_d$ = 595 nm \\
$l_2$ = 1.53 $\mu$m~~~~~~ & $w_2$ = 177 nm~~~~~~ & $R_r$ = 5.26 $\mu$m  \\
$l_3$ = 4.20 $\mu$m~~~~~~ & $w_3$ = 151 nm~~~~~~ & $\theta$ = 92.7 deg \\
\hline
\end{tabular}
\caption{Numerical values for the dimensions of the device studied in this work. Measurements were performed using the scanning electron microscope image shown in Fig.~\ref{fig2}(a). Device thickness was taken to be $d$ = 250 nm as specified by the manufacturer.}
\label{tab1}
\end{table}

\section{Cavity Optomechanics with Radiation Pressure and Photothermal Effects}
\label{omeoms}

In this section, we look to theoretically model the behavior of an optomechanical cavity that is subject to both radiation pressure and photothermal effects. Specifically, we look to determine in which parameter regimes these individual forces will dominate each of the optomechanical damping and spring effect, as well as investigate how the inclusion of the photothermal force modifies the radiation-pressure-driven optomechanical phenomena of cooling and nonlinear parametric amplification.

To begin, we consider an optical cavity that is dispersively coupled to the position of a mechanical resonator in the presence of both radiation pressure and photothermal optical forces. Treating the system semiclassically, the equations of motion for the field amplitude $a$ of the cavity and mechanical position $x$ of the resonator will be given by \cite{pinard_2008, restrepo_2011, deliberato_2011, abdi_2012a, abdi_2012b, aspelmeyer_2014, harris_2016}
\begin{gather}
\begin{gathered}
\label{aeom_app}
\dot{a}(t) = -\frac{\kappa}{2} a(t) + i \Delta_0 a(t) + i G x(t) a(t) + \sqrt{\kappa_{\rm e}} a_{\rm in}(t) \\
+ \sqrt{\kappa_{\rm a}} a_{\rm abs}'(t) + \sqrt{\kappa_{\rm o}} a_{\rm o}'(t), 
\end{gathered} \\
\label{xeom_app}
\ddot{x}(t) + \Gamma_{\rm m} \dot{x}(t) + \omega_{\rm m}^2 x(t) = \frac{1}{m} \left[ F_{\rm th}(t) + F_{\rm rp}(t) + F_{\rm pt}(t) \right].
\end{gather}
Here, $\Delta_0 = \omega_{\rm d} - \omega_{\rm c}$ is the bare detuning of the optical drive frequency $\omega_{\rm d}$ from the resonant frequency of the cavity $\omega_{\rm c}$, while $\kappa = \kappa_{\rm e} + \kappa_{\rm a} + \kappa_{\rm o}$ is the total decay rate of the optical cavity, comprised of contributions from losses to the external coupler, absorption of cavity photons in the mechanical element and all other sources, at the associated rates of $\kappa_{\rm e}$, $\kappa_{\rm a}$ and $\kappa_{\rm o}$, respectively. Note that both $\kappa_{\rm a}$ and $\kappa_{\rm o}$ contribute to the total intrinsic damping rate of the optical cavity $\kappa_{\rm i} = \kappa_{\rm a} + \kappa_{\rm o}$. We then have that $a_{\rm in}(t)$ is the field input to the cavity via the external coupler, while $a_{\rm abs}'(t)$ and $a_{\rm o}'(t)$ are the field operators associated with vacuum noise (denoted by primes) that creeps into the system via absorption of photons in the mechanical element and other loss channels, respectively, with each of these input operators having units of $1/\sqrt{\rm s}$. Meanwhile, $\omega_{\rm m}$, $\Gamma_{\rm m}$, and $m$ are the resonant frequency, damping rate, and effective mass of the mechanical resonator, which is actuated by an intrinsic thermal fluctuation force $F_{\rm th}$, as well as two optically-driven forces, the radiation pressure force \cite{aspelmeyer_2014, harris_2016}
\begin{equation}
\label{Frp}
F_{\rm rp}(t) = \hbar G a^\dag (t) a(t),
\end{equation}
and the photothermal force \cite{pinard_2008, restrepo_2011, abdi_2012a, abdi_2012b}
\begin{equation}
\label{Fpt_app}
F_{\rm pt}(t) = \frac{\hbar G \beta}{\kappa_{\rm a} \tau} \int_{-\infty}^t e^{-\frac{t-t'}{\tau}} a_{\rm abs}^\dag(t') a_{\rm abs}^{\phantom \dag}(t') dt'.
\end{equation}
Here, $G = - d \omega_{\rm c} / dx$ is the optomechanical coupling coefficient and $a_{\rm abs}(t) = \sqrt{\kappa_a} a(t) - a_{\rm abs}'(t)$ is the field operator characterizing the photons absorbed by the mechanical resonator \cite{pinard_2008, restrepo_2011, abdi_2012a}. Also included in Eq.~\eqref{Fpt_app} is the photothermal time constant $\tau$ that sets the timescale of the photothermal force (see Appendix \ref{tauthsec} for more details) and the dimensionless parameter $\beta$ that determines the relative strength and direction of the photothermal force with respective to the radiation pressure force. It is important to note that it is possible for $\beta$ to be negative in value \cite{restrepo_2011, metzger_2004, jourdan_2008, yuvaraj_2012}, such that the photothermal force acts to directly oppose radiation pressure effects. Finally, while we have chosen to identify the force in Eq.~\eqref{Fpt_app} as photothermal in nature, with the appropriate choice of $\tau$, $\beta$ and $\kappa_{\rm a}$ the description that follows can be applied to any optical force that is delayed in time.

\subsection{Linearized Optomechanical Equations of Motion}
\label{linom}

In order to solve the equations of motion for the above optomechanical system, we express each quantity as a combination of its classical, steady-state amplitude (denoted by an overhead bar) and its fluctuations about this mean value (denoted by an operator hat). This leads to $a(t) = \bar{a} + \hat{a}(t)$, $a_{\rm in}(t) = \bar{a}_{\rm in} + \hat{a}'_{\rm in}(t)$, $a_{\rm abs}(t) = \bar{a}_{\rm abs} + \hat{a}_{\rm abs}(t)$, $a'_{\rm abs}(t) = \hat{a}'_{\rm abs}(t)$, $a'_{\rm o}(t) = \hat{a}'_{\rm o}(t)$, $x(t) = \bar{x} + \hat{x}(t)$, and $F_{\rm th}(t) = \hat{F}_{\rm th}(t)$. Note that each of the noise quantities [{\it i.e.}, $a'_{\rm abs}(t)$, $a'_{\rm o}(t)$ and $F_{\rm th}(t)$] are comprised solely of a fluctuation term (which includes both thermal and quantum noise). Inputting each of these relations into Eqs.~\eqref{aeom_app}--\eqref{Fpt_app}, while only keeping terms to first order in the fluctuations, we linearize Eqs.~\eqref{aeom_app} and \eqref{xeom_app}, resulting in
\begin{gather}
\begin{gathered}
\label{ahat}
\dot{\hat{a}}(t) = -\frac{\kappa}{2} \hat{a}(t) + i \Delta \hat{a}(t) + i G \bar{a} \hat{x}(t) + \sqrt{\kappa_{\rm e}} \hat{a}'_{\rm in}(t) \\
+ \sqrt{\kappa_{\rm a}} \hat{a}'_{\rm abs}(t) + \sqrt{\kappa_{\rm o}} \hat{a}'_{\rm o}(t), 
\end{gathered} \\
\begin{gathered}
\label{xhat}
\ddot{\hat{x}}(t) + \Gamma_{\rm m} \dot{\hat{x}}(t) + \omega_{\rm m}^2 \hat{x}(t) = \frac{1}{m} \bigg[ \hat{F}_{\rm th}(t) + \hbar G \bar{a} \left[ \hat{a}(t) + \hat{a}^\dag(t) \right] \\ 
+ \frac{\hbar G \beta \bar{a}}{\tau} \int_{-\infty}^t e^{-\frac{t-t'}{\tau}} \bigg( \left[ \hat{a}(t') + \hat{a}^\dag(t') \right] \\ 
- \frac{1}{\sqrt{\kappa_{\rm a}}} \left[ \hat{a}'_{\rm abs}(t') + \hat{a}'^\dag_{\rm abs}(t') \right] \bigg) dt \bigg],
\end{gathered}
\end{gather}
with the steady-state values of $a$ and $x$ being
\begin{gather}
\label{abar}
\bar{a} = \frac{\sqrt{\kappa_{\rm e}} \bar{a}_{\rm in}}{\kappa / 2 - i \Delta}, \\
\label{xbar}
\bar{x} = \frac{\hbar G |\bar{a}|^2 (1 + \beta)}{m \omega_{\rm m}^2}.
\end{gather}
Note that we have introduced a new cavity detuning $\Delta = \Delta_0 + G \bar{x}$ to account for the static shift in cavity frequency due to the steady-state displacement of the mechanical equilibrium position. 

In this linearized form, Eqs.~\eqref{ahat} and \eqref{xhat} can now be Fourier transformed, resulting in the frequency representation of the cavity field and mechanical displacement fluctuations as
\begin{gather}
\begin{gathered}
\label{ahatw}
\hat{a}(\omega) = \chi_{\rm c}(\omega) \big[ i G \bar{a} \hat{x}(\omega) + \sqrt{\kappa_{\rm e}} \hat{a}'_{\rm in}(\omega) \\ 
+ \sqrt{\kappa_{\rm a}} \hat{a}'_{\rm abs}(\omega) + \sqrt{\kappa_{\rm o}} \hat{a}'_{\rm o}(\omega) \big],
\end{gathered} \\
\begin{gathered}
\label{xhatw}
\hat{x}(\omega) = \chi_{\rm m}(\omega) \bigg[ \hat{F}_{\rm th}(\omega) + \hbar G \bar{a} \bigg\{ \left( 1 + \frac{\beta}{1 - i \omega \tau} \right) \left[\hat{a}(\omega) + \hat{a}^\dag(\omega) \right] \\ 
- \frac{\beta}{\sqrt{\kappa_{\rm a}}(1 - i \omega \tau)} \left[ \hat{a}'_{\rm abs}(\omega) + \hat{a}'^\dag_{\rm abs}(\omega) \right] \bigg\} \bigg],
\end{gathered}
\end{gather}
where we have introduced the frequency-dependent susceptibilities of the optical cavity $\chi_{\rm c}(\omega)$ and mechanical resonator $\chi_{\rm m}(\omega)$ as
\begin{gather}
\label{chic}
\chi_{\rm c}(\omega) = \frac{1}{\kappa / 2 - i (\Delta + \omega)}, \\
\label{chim}
\chi_{\rm m}(\omega) = \frac{1}{m \left(\omega_{\rm m}^2 - \omega^2 - i \omega \Gamma_{\rm m} \right)}.
\end{gather}

\subsection{Optomechanical Damping and Spring Effect}
\label{omwmGm}

To determine the optomechanical damping and spring effect, we input Eq.~\eqref{ahatw} into Eq.~\eqref{xhatw}, resulting in
\begin{equation}
\begin{split}
\label{xhatwcoup}
\hat{x}(\omega) &= \chi_{\rm eff}(\omega) \bigg[ \hat{F}_{\rm th}(\omega) + \hbar G \bar{a} \bigg\{ \left( 1 + \frac{\beta}{1 - i \omega \tau} \right) \\
&\times \bigg( \chi_{\rm c}(\omega) \left[\sqrt{\kappa_{\rm e}} \hat{a}'_{\rm in}(\omega) + \sqrt{\kappa_{\rm a}} \hat{a}'_{\rm abs}(\omega) + \sqrt{\kappa_{\rm o}} \hat{a}'_{\rm o}(\omega) \right] \\ 
&+ \chi^*_{\rm c}(-\omega) \left[ \sqrt{\kappa_{\rm e}} \hat{a}'^\dag_{\rm in}(\omega) + \sqrt{\kappa_{\rm a}} \hat{a}'^\dag_{\rm abs}(\omega) + \sqrt{\kappa_{\rm o}} \hat{a}'^\dag_{\rm o}(\omega) \right] \bigg) \\ 
&- \frac{\beta}{\sqrt{\kappa_{\rm a}}(1 - i \omega \tau)} \left[ \hat{a}'_{\rm abs}(\omega) + \hat{a}'^\dag_{\rm abs}(\omega) \right] \bigg\} \bigg],
\end{split}
\end{equation}
where we have introduced the effective mechanical susceptibility defined as \cite{pinard_2008, restrepo_2011, abdi_2012a, harris_2016}
\begin{equation}
\begin{split}
\label{chieff}
\chi^{-1}_{\rm eff}(\omega) &= \chi^{-1}_{\rm m}(\omega) \\ 
- i \hbar & G^2 |\bar{a}|^2 \left( 1 + \frac{\beta}{1 - i \omega \tau} \right)  \bigg[\chi_{\rm c}(\omega) - \chi_{\rm c}^*(-\omega) \bigg] \\
&\equiv m \left( \omega_{\rm m}^2 - \omega^2 + 2 \omega \delta \omega_{\rm m} - i \omega  \Gamma_{\rm m} + \delta \Gamma_{\rm m} \right).
\end{split}
\end{equation}
From this effective susceptibility, we can extract the optomechanically induced shift in the mechanical resonance frequency, or optomechanical spring effect, 
\begin{equation}
\begin{split}
\label{dwm}
&\delta \omega_{\rm m} = -\frac{\hbar G^2 |\bar{a}|^2}{2 m \omega_{\rm m}}{\rm Re} \left\{ i \left( 1 + \frac{\beta}{1 - i \omega_{\rm m} \tau} \right) \bigg[\chi_{\rm c}(\omega) - \chi_{\rm c}^*(-\omega) \bigg] \right\} \\
&= 2 \bar{N} g_0^2 \Delta |\chi_{\rm c} (\omega_{\rm m})|^2 |\chi_{\rm c}(-\omega_{\rm m})|^2 \\
&\times \left[ \frac{\kappa^2}{4} + \Delta^2 -\omega_{\rm m}^2 + \frac{\beta}{1 + \omega_{\rm m}^2 \tau^2} \left( \frac{\kappa^2}{4} + \Delta^2 - \omega_{\rm m}^2 - \omega_{\rm m}^2 \kappa \tau \right) \right],
\end{split}
\end{equation}
as well as the shift in the mechanical damping rate, or optomechanical damping, 
\begin{equation}
\begin{split}
\label{dGm}
&\delta \Gamma_{\rm m} = \frac{\hbar G^2 |\bar{a}|^2}{m \omega_{\rm m}}{\rm Im} \left\{i \left( 1 + \frac{\beta}{1 - i \omega_{\rm m} \tau} \right) \bigg[\chi_{\rm c}(\omega) - \chi_{\rm c}^*(-\omega) \bigg] \right\} \\
&= - 4 \bar{N} g_0^2 \Delta \omega_{\rm m} |\chi_{\rm c} (\omega_{\rm m})|^2 |\chi_{\rm c}(-\omega_{\rm m})|^2 \\
&\times \left\{ \kappa + \frac{\beta}{1 + \omega_{\rm m}^2 \tau^2} \left[ \kappa + \tau \left( \frac{\kappa^2}{4} + \Delta^2 -\omega_{\rm m}^2 \right) \right] \right\}.
\end{split}
\end{equation}
Here $\bar{N} = |\bar{a}|^2$ is the average number of coherent photons confined to the optical cavity, while $g_0 = G x_{\rm zpf}$ is the single-phonon, single-phonon optomechanical coupling rate, with $x_{\rm zpf} = \sqrt{\hbar / 2 m \omega_{\rm m}}$ being the zero-point fluctuation amplitude of the mechanical resonator. Note that we have also taken $\omega \approx \omega_{\rm m}$ in Eqs.~\eqref{dwm} and \eqref{dGm}, as we are only concerned with effects near mechanical resonance.  As expected, these dynamical backaction effects vanish for zero detuning ($\Delta = 0$) and the standard radiation-pressure-driven expressions are restored when $\beta = 0$ \cite{aspelmeyer_2014}.

We are now interested in determining the parameter space for which the optomechanical damping and spring effect are dominated by photothermal forces. By inspection of Eqs.~\eqref{dwm} and \eqref{dGm}, we find that this will occur for the spring effect when
\begin{equation}
\label{dwmptdom}
\left| \frac{\kappa^2}{4} + \Delta^2 -\omega_{\rm m}^2 \right| < \left| \frac{\beta}{1 + \omega_{\rm m}^2 \tau^2} \left( \frac{\kappa^2}{4} + \Delta^2 - \omega_{\rm m}^2 - \omega_{\rm m}^2 \kappa \tau \right) \right|,
\end{equation}
while the photothermal force will dominate the optomechanical damping if
\begin{equation}
\label{dGmptdom}
\kappa < \left| \frac{\beta}{1 + \omega_{\rm m}^2 \tau^2} \left[ \kappa + \tau \left( \frac{\kappa^2}{4} + \Delta^2 -\omega_{\rm m}^2 \right) \right] \right|.
\end{equation}
These inequalities are simplified considerably if we restrict ourselves to the experimentally relevant parameter space of $\kappa \gg \omega_{\rm m}$ ({\it i.e.}, the non-sideband-resolved regime) and $\omega_{\rm m} \tau \sim 1$, which together imply $\kappa \tau \gg 1$. Using these conditions, Eq.~\eqref{dwmptdom} will be satisfied if
\begin{equation}
\label{dwmptdomfin_app}
1 + \omega_{\rm m}^2 \tau^2 \lesssim |\beta|,
\end{equation}
while Eq.~\eqref{dGmptdom} becomes
\begin{equation}
\label{dGmptdomsim}
\kappa < \frac{\left| \beta \right| \tau}{1 + \omega_{\rm m}^2 \tau^2} \left( \frac{\kappa^2}{4} + \Delta^2 \right).
\end{equation}
Finally, Eq.~\eqref{dGmptdomsim} can be further simplified if we assume $\Delta \sim \pm \kappa/2$, {\it i.e.}, only consider the region where optomechanical damping is maximized, which results in
\begin{equation}
\label{dGmptdomfin_app}
1 + \omega_{\rm m}^2 \tau^2 \lesssim \frac{\left| \beta \right| \kappa \tau}{2}.
\end{equation}
Note that for the above inequalities we have taken the absolute value of $\beta$ as it can be positive or negative depending on the orientation of the photothermal force with respect to the radiation pressure force. From Eqs.~\eqref{dwmptdomfin_app}--\eqref{dGmptdomfin_app} it is therefore clear that for $\kappa \tau \gg 1$ (as is assumed here and is experimentally relevant for this work), it is possible to have values of $\beta$ such that radiation pressure forces dominate the spring effect, while photothermal effects dictate the optomechanical damping. We note that this is especially important in the non-sideband-resolved regime, where $\kappa$ is generally large, as highlighted by the fact that photothermal damping effects are often stronger in non-sideband-resolved cavities when compared to their sideband-resolved counterparts \cite{pinard_2008, restrepo_2011, abdi_2012a, abdi_2012b}. Furthermore, if $\beta < 0$, the photothermal and radiation pressure forces oppose each other, resulting in an oddly similar detuning dependence between the optomechanical spring effect and damping in apparent violation of the Kramers-Kronig relations \cite{aspelmeyer_2014b} [see Fig.~\ref{fig1}(c)/(f)].

\subsection{Optomechanical Cooling}
\label{omcool}

We now look to see how the inclusion of a photothermal force acts to modify conventional radiation-pressure-driven backaction cooling. To do this, we begin by determining the two-sided spectral density of the mechanical displacement $S_{xx}(\omega)$ in the presence of optomechanical effects, which can be found by using \cite{hauer_2015}
\begin{equation}
\label{Sxxdef}
S_{xx}(\omega) = \frac{1}{2 \pi} \int_{-\infty}^{\infty} \braket{\hat{x}(\omega) \hat{x}(\omega')} d \omega',
\end{equation}
along with the following Markovian noise correlators \cite{pinard_2008, aspelmeyer_2014}
\begin{align}
\label{Fthcorr}
\braket{\hat{F}_{\rm th}(\omega) \hat{F}_{\rm th}(\omega')} &= 2 \pi \hbar \omega m \Gamma_{\rm m} \coth \left( \frac{\hbar \omega}{2 k_{\rm B} T} \right) \delta(\omega + \omega'), \\
\label{acorr1}
\braket{\hat{a}'_i(\omega) \hat{a}'^\dag_i(\omega')} &= 2 \pi \delta (\omega + \omega'), \\
\label{acorr2}
\braket{\hat{a}'_i(\omega) \hat{a}'_i(\omega')} &= \braket{\hat{a}'^\dag_i(\omega) \hat{a}'_i(\omega')} = \braket{\hat{a}'^\dag_i(\omega) \hat{a}'^\dag_i(\omega')} = 0.
\end{align}
Note that in Eqs.~\eqref{acorr1} and \eqref{acorr2} we have used $\hat{a}'_i$ as a placeholder for any of the optical vacuum fluctuation amplitudes $\hat{a}'_{\rm in}$, $a'_{\rm abs}$, and $a'_{\rm o}$, as well as assumed a zero temperature bath for each optical mode (due to the fact that $\hbar \omega_{\rm c} \gg k_{\rm B} T$). Inputting Eqs.~\eqref{xhatw} and \eqref{Fthcorr}--\eqref{acorr2} into Eq.~\eqref{Sxxdef}, while using the fact that $\chi_{\rm eff}(-\omega) = \chi_{\rm eff}^*(\omega)$ [this is a direct consequence of $\chi_{\rm eff}(t)$ being a real-valued function], we then find
\begin{equation}
\label{Sxx}
S_{xx}(\omega) = | \chi_{\rm eff}(\omega)|^2 \left[ S_{FF}^{\rm th}(\omega) + S_{FF}^{\rm opt} (\omega) \right],
\end{equation}
where
\begin{equation}
\begin{split}
\label{SFFth}
S_{FF}^{\rm th}(\omega) &= \frac{1}{2 \pi} \int_{-\infty}^{\infty} \braket{\hat{F}_{\rm th}(\omega) \hat{F}_{\rm th}(\omega')} d \omega' \\ 
&= \hbar \omega m \Gamma_{\rm m} \coth \left( \frac{\hbar \omega}{2 k_{\rm B} T} \right),
\end{split}
\end{equation}
is the spectral density of the thermal force \cite{hauer_2015, pinard_2008} and $S_{FF}^{\rm opt}(\omega) = S_{FF}^{\rm rp}(\omega) + S_{FF}^{\rm pt}(\omega)$ is the optical force spectral density, comprised of the spectra due to radiation pressure $S_{FF}^{\rm rp}(\omega)$ and photothermal effects $S_{FF}^{\rm pt}(\omega)$. We find it convenient to express these spectra as $S_{FF}^{\rm rp}(\omega) = \hbar^2 G^2 S_{NN}^{\rm rp}(\omega)$ and $S_{FF}^{\rm pt}(\omega) = \hbar^2 G^2 S_{NN}^{\rm pt}(\omega)$ where
\begin{equation}
\label{SNNrp}
S_{NN}^{\rm rp}(\omega) =  \frac{\bar{N} \kappa}{\left( \omega - \Delta \right)^2 + \left( \kappa / 2 \right)^2},
\end{equation}
and
\begin{equation}
\begin{split}
\label{SNNpt}
&S_{NN}^{\rm pt}(\omega) =  \frac{\bar{N} }{\left( \omega - \Delta \right)^2 + \left( \kappa / 2 \right)^2} \frac{\beta}{1 + \omega_{\rm m}^2 \tau^2} \\
&\times \left[ \kappa \left( \frac{\beta \kappa}{4 \kappa_{\rm a}} + 1 \right) + \left(\Delta + \omega \right) \left( \frac{\beta (\Delta + \omega)}{\kappa_{\rm a}} - 2 \omega \tau \right) \right],
\end{split}
\end{equation}
are the effective cavity photon number spectra associated with the radiation pressure and photothermal forces, respectively \cite{marquardt_2007, clerk_2010}.

Using the spectral density function given by Eq.~\eqref{Sxx}, we can determine the mean-squared value of the mechanical displacement using the relation \cite{hauer_2015}
\begin{equation}
\label{x2meandef}
\braket{x^2} = \frac{1}{2 \pi} \int_{-\infty}^{\infty} S_{xx}(\omega) d \omega.
\end{equation}
To perform this integral, we use the approximation
\begin{equation}
\label{chieffapp}
|\chi_{\rm eff}(\omega)|^2 \approx \frac{\pi}{2 m^2 \omega_{\rm m}^2 \Gamma_{\rm tot}} \left[ \delta (\omega - \omega_{\rm m}) + \delta ( \omega + \omega_{\rm m}) \right],
\end{equation}
where $\Gamma_{\rm tot} = \Gamma_{\rm m} + \delta \Gamma_{\rm m}$ is the total mechanical damping rate, including both the intrinsic mechanical damping and optomechanical effects. This approximation is valid for a high-$Q$ mechanical resonator ({\it i.e.}, $Q_{\rm m} = \omega_{\rm m} / \Gamma_{\rm m} \gg 1$), due to the fact that the majority of the mechanical displacement spectrum is located near $\omega \approx \pm \omega_{\rm m}$. Using this approximation to evaluate the integral in Eq.~\eqref{x2meandef}, we find
\begin{equation}
\label{x2mean}
\braket{x^2} = \frac{x_{\rm zpf}^2}{\Gamma_{\rm tot}} \left\{ ( 2 \bar{n}_{\rm th} + 1) \Gamma_{\rm m} + g_0^2 \left[ S_{NN}^{\rm opt}(\omega_{\rm m}) + S_{NN}^{\rm opt}(-\omega_{\rm m}) \right] \right\},
\end{equation}
where we have taken advantage of the relation $\coth \left( \hbar \omega/2 k_{\rm B} T \right) = 2 \bar{n}_{\rm th} + 1$, with $\bar{n}_{\rm th} = \left( e^{\hbar \omega_{\rm m} / k_B T} - 1 \right)^{-1}$ being the average thermal phonon occupation number of the bath according to Bose-Einstein statistics evaluated at the mechanical resonance frequency. Comparing Eq.~\eqref{x2mean} to the expected expression for the mean-squared displacement of the resonator, $\braket{x^2} = 2 x_{\rm zpf} \left(\braket{n} + \frac{1}{2} \right)$ \cite{hauer_2015}, we determine the average phonon occupancy $\braket{n}$ of a mechanical resonator subject to both photothermal and radiation pressure optomechanical forces to be
\begin{equation}
\label{nave}
\braket{n} = \frac{( 2 \bar{n}_{\rm th} + 1) \Gamma_{\rm m} + g_0^2 \left[ S_{NN}^{\rm opt}(\omega_{\rm m}) + S_{NN}^{\rm opt}(-\omega_{\rm m}) \right]}{2 \Gamma_{\rm tot}} - \frac{1}{2}.
\end{equation}
Finally, we note that using the identity \cite{marquardt_2007, clerk_2010}
\begin{equation}
\begin{split}
\label{delGamid}
\delta \Gamma_{\rm m} &= \frac{x_{\rm zpf}^2}{\hbar^2} \left[ S_{FF}^{\rm opt}(\omega_{\rm m}) - S_{FF}^{\rm opt}(-\omega_{\rm m}) \right] \\
 &= g_0^2 \left[ S_{NN}^{\rm opt}(\omega_{\rm m}) - S_{NN}^{\rm opt}(-\omega_{\rm m}) \right],
\end{split}
\end{equation}
we can recast Eq.~\eqref{nave} into the familiar rate equation form \cite{marquardt_2007, restrepo_2011, clerk_2010}
\begin{equation}
\label{naverate}
\braket{n} = \frac{\bar{n}_{\rm th} \Gamma_{\rm m} + \bar{n}_{\rm min} \delta \Gamma_{\rm m}}{\Gamma_{\rm tot}},
\end{equation}
allowing us to identify the minimum attainable average phonon occupancy using this cooling method as 
\begin{equation}
\begin{split}
\label{nmin_app}
&\bar{n}_{\rm min} = \left[ S_{NN}^{\rm opt}(\omega_{\rm m})/S_{NN}^{\rm opt}(-\omega_{\rm m}) - 1 \right]^{-1} \\
&= -\frac{ \frac{\kappa^2}{4} + (\Delta + \omega_{\rm m})^2 }{4 \Delta \omega_{\rm m} \left\{\kappa + \frac{\beta}{1 + \omega_{\rm m}^2 \tau^2} \left[ \kappa + \tau \left( \frac{\kappa^2}{4} + \Delta^2 -\omega_{\rm m}^2 \right) \right] \right\}} \\
&\times \bigg\{ \kappa +  \frac{\beta}{1 + \omega_{\rm m}^2 \tau^2} \bigg[ \kappa \left( \frac{\beta \kappa}{4 \kappa_{\rm a}} + 1 \right) \\
&+ \left(\Delta + \omega_{\rm m} \right) \left( \frac{\beta (\Delta + \omega_{\rm m})}{\kappa_{\rm a}} - 2 \omega_{\rm m} \tau \right) \bigg] \bigg\}.
\end{split}
\end{equation}
As expected, by setting $\beta = 0$, Eq.~\eqref{nmin_app} reverts to the standard radiation pressure result \cite{aspelmeyer_2014, marquardt_2007}
\begin{equation}
\label{nminrp_app}
\bar{n}_{\rm min}^{\rm rp} = - \frac{\frac{\kappa^2}{4} + (\Delta + \omega_{\rm m})^2 }{ 4 \Delta \omega_{\rm m}}.
\end{equation}
Furthermore, if we sever the connection to the optomechanical bath [{\it i.e.}, set $G = g_0 = 0$ in Eqs.~\eqref{nave} and \eqref{naverate}], the mechanical resonator thermalizes to its environmental bath such that $\braket{n} = \bar{n}_{\rm th}$.

The quantity in Eq.~\eqref{nmin_app} describes the minimum attainable phonon occupation of the mechanical resonator, which can be reached if $\delta \Gamma_{\rm m}$ is large enough such that $\Gamma_{\rm tot} \approx \delta \Gamma_{\rm m}$ and $\bar{n}_{\rm th} \Gamma_{\rm m} \ll \bar{n}_{\rm min} \delta \Gamma_{\rm m}$. In this sense, Eqs.~\eqref{nave} and \eqref{naverate} do not include effects that would arise when experimentally performing optomechanical cooling of the mechanical resonator, such as the inevitable heating due to photon absorption \cite{deliberato_2011}. Nevertheless, the inclusion of photothermal effects has a substantial influence on this minimal phonon occupation when compared to the result obtained using solely radiation pressure, especially in the non-sideband-resolved regime \cite{pinard_2008, deliberato_2011, restrepo_2011, abdi_2012a, abdi_2012b}. In fact, due to interference between the radiation pressure and photothermal forces, it is possible to cool the mechanical resonator to an average phonon occupancy below one while operating in the non-sideband-resolved regime \cite{pinard_2008, deliberato_2011, restrepo_2011, abdi_2012a, abdi_2012b}, a feat which is not possible for a dispersively coupled, radiation-pressure-driven optomechanical cavity \cite{aspelmeyer_2014, marquardt_2007}.

\begin{figure}[h!]
\centerline{\includegraphics[width=3.0in]{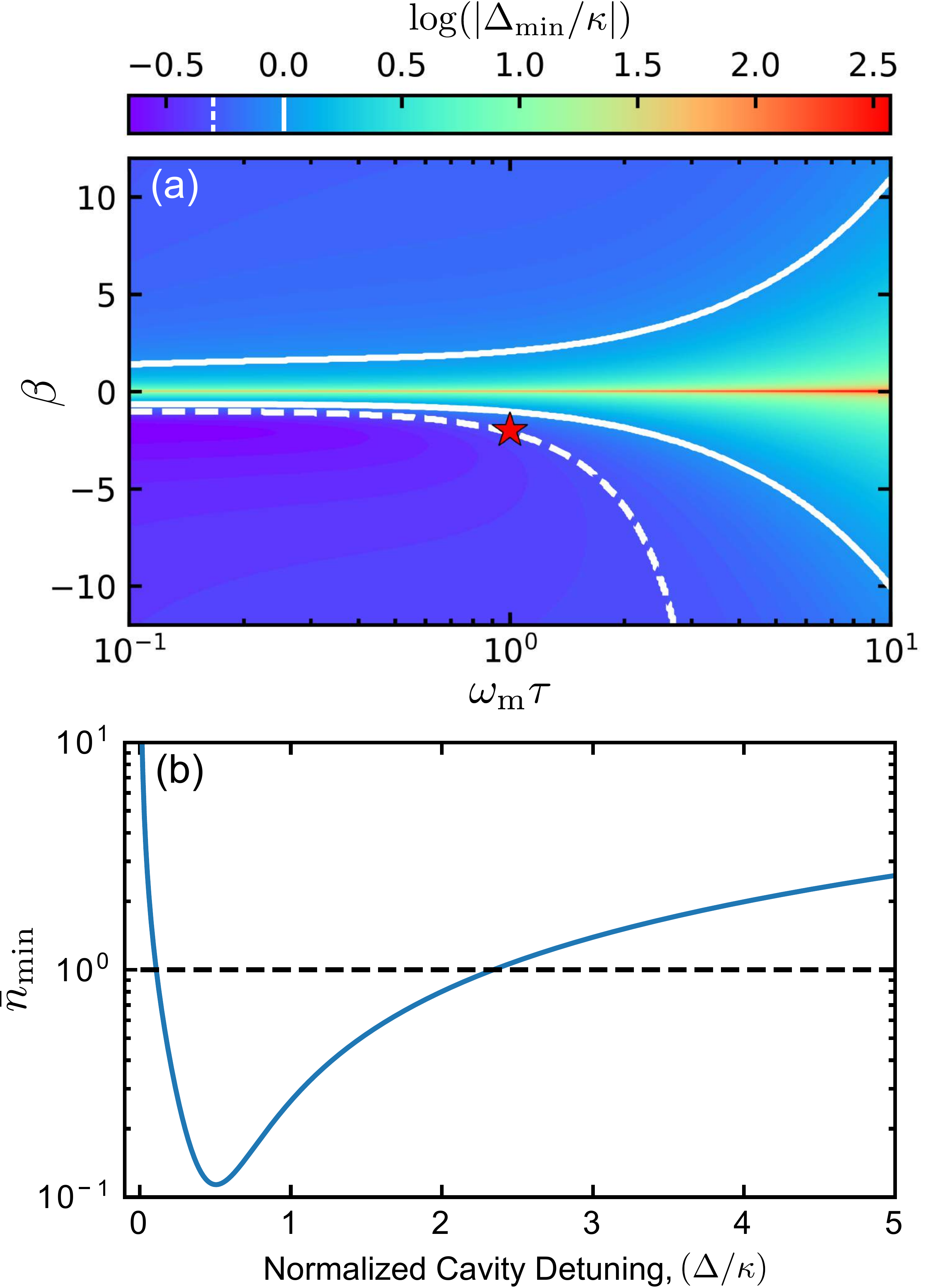}}
\caption{{\label{fig9}} (a) Color plot of ${\rm log}(|\Delta_{\rm min} / \kappa|)$, {\it i.e.}, the base-ten logarithm of the absolute value of the detuning $\Delta_{\rm min}$ for which $\bar{n}_{\rm min}$ is minimized (normalized by $\kappa$). Here we have used the same optomechanical parameters as those in Fig.~\ref{fig6}, with the solid (dashed) white line indicating the contour of $\Delta_{\rm min} / \kappa = 1.0$ ($\Delta_{\rm min} / \kappa = 0.5$). As one can see, $\Delta_{\rm min}$ is maximized near $\beta = 0$ and increases with larger values of $\omega_{\rm m} \tau$. Furthermore, as is seen with $\bar{n}_{\rm min}$ in Fig.~\ref{fig6}(b), $\Delta_{\rm min}$ is asymmetric with respect to $\beta$, which is again due to interference between radiation pressure and photothermal effects. We note that the dashed contour of $\Delta_{\rm min} / \kappa = 0.5$ passes very near $\beta = -2.0$ for $\omega_{\rm m} \tau = 1$ (see red star), such that at this point the photothermal cooling is optimized versus both cavity detuning and thermal relaxation time. In (b), we plot $\bar{n}_{\rm min}$ versus detuning for this special set of parameters, showing that the lowest achievable phonon number is indeed minimized to $\bar{n}_{\rm min} = 0.11$ at $\Delta_{\rm min} \approx 0.5 \kappa$.}
\end{figure}

In order to determine the absolute minimum phonon occupancy that can be reached for a given optomechanical cavity, the expressions in Eqs.~\eqref{nmin_app} and \eqref{nminrp_app} must be minimized with respect to the cavity detuning $\Delta$. This is easily done for the case of radiation pressure alone, where Eq.~\eqref{nminrp_app} is found to be minimized for $\Delta_{\rm min}^{\rm rp} = \sqrt{\kappa^2/4 + \omega_{\rm m}^2}$. Therefore, in the non-sideband-resolved regime, $\Delta_{\rm min}^{\rm rp} \approx \kappa / 2$, leading to $\bar{n}_{\rm min}^{\rm rp} \approx \kappa / 4 \omega_{\rm m}$ \cite{aspelmeyer_2014}. Unfortunately, the situation is far more complicated when the photothermal force is included, such that we are unable to determine a closed-form solution for the detuning $\Delta_{\rm min}$ that minimizes Eq.~\eqref{nmin_app}. However, $\Delta_{\rm min}$ can be determined numerically for a given set of conditions, as we have shown in Fig.~\ref{fig9}(a) for the same parameter space that is mapped out in Fig.~\ref{fig6}(b). Here we see that $\Delta_{\rm min}$ grows for decreasing $\beta$ and increasing $\omega_{\rm m} \tau$, moving away from the optimal value of $\Delta_{\rm min} / \kappa = 0.5$ denoted by the white dashed line, while exhibiting a similar asymmetry about $\beta$ as was seen for $\bar{n}_{\rm min}$. Interestingly, this plot further shows that near $\beta = -2.0$ and $\omega_{\rm m} \tau = 1$, we find $\Delta_{\rm min} / \kappa \approx 0.5$ [see red star in Fig.~\ref{fig9}(a)], such that for this set of parameters the strength of the photothermal force is maximized with respect to both detuning and thermal relaxation time \cite{deliberato_2011, metzger_2004, metzger_2008a}, while still allowing for ground state cooling of the mechanical motion to an occupancy as low as $n_{\rm min} = 0.11$ [see Fig.~\ref{fig9}(b)].

\subsection{Nonlinear Optomechanics}
\label{nlom}

In the previous sections of this appendix, we implicitly assumed an optomechanical system whose mechanical fluctuations are small enough to allow for a linearized treatment of the equations of motion. However, when the amplitude of oscillation $A$ of the mechanical resonator becomes large enough ($GA \gg \omega_{\rm m}$), it is possible to enter into a regime where keeping terms to first order in their fluctuations no longer suffices. One such situation where this occurs is optomechanical self-amplification \cite{marquardt_2006, ludwig_2008, metzger_2008b, aspelmeyer_2014, krause_2015}, which onsets when $\delta \Gamma_{\rm m} = - \Gamma_{\rm m}$, such that $\Gamma_{\rm tot}$ drops to zero and a parametric instability emerges, driving the mechanical motion into large amplitude oscillations in order to counteract the optical drive forces.

Here we will study this nonlinear optomechanical interaction in the same context as the previous sections, where we include both radiation pressure and photothermal effects. However, as the large amplitude mechanical oscillations associated with this nonlinear regime, along with the cavity drive field, act to overwhelm any quantum noise [{\it i.e.}, terms containing $a'_{\rm in}(t)$, $a'_{\rm abs}(t)$, or $a'_{\rm o}(t)$], we restrict ourselves to a classical treatment of the optomechanical system, such that Eqs.~\eqref{aeom_app} and \eqref{xeom_app} become
\begin{gather}
\label{aeomcl}
\dot{a}(t) = -\frac{\kappa}{2} a(t) + i \Delta_0 a(t) + i G x(t) a(t) + \sqrt{\kappa_{\rm e}} \bar{a}_{\rm in}, \\
\label{xeomcl}
\ddot{x}(t) + \Gamma_{\rm m} \dot{x}(t) + \omega_{\rm m}^2 x(t) = \frac{1}{m} \left[F_{\rm th}(t) + F_{\rm rp}(t) + F_{\rm pt}(t) \right].
\end{gather}
Furthermore, we introduce the classical photothermal force by modifying Eq.~\eqref{Fpt_app} to obtain
\begin{equation}
F_{\rm pt}(t) = \frac{\hbar G \beta}{\tau} \int_{-\infty}^t e^{-\frac{t-t'}{\tau}} a^\dag(t') a(t') dt'.
\label{Fptcl}
\end{equation}
We continue by assuming a high-$Q$ mechanical system, such that we can use the ansatz \cite{marquardt_2006, ludwig_2008, aspelmeyer_2014}
\begin{equation}
\label{xoft}
x(t) = \bar{x} + A \cos(\omega_{\rm m} t),
\end{equation}
as the solution to Eq.~\eqref{xeomcl} for the resonator's displacement, where again $\bar{x}$ is the resonator's static displacement from equilibrium. Inputting this expression into Eq.~\eqref{aeomcl}, we solve for the optical field amplitude as \cite{marquardt_2006, aspelmeyer_2014, krause_2015}
\begin{equation}
\label{aoft_app}
a(t) = \sqrt{\kappa_{\rm e}} \bar{a}_{\rm in} e^{i \phi(t)} \sum_{k=-\infty}^\infty \alpha_k e^{i k \omega_{\rm m} t},
\end{equation}
where $\phi(t) = \xi \sin (\omega_{\rm m} t)$ is a time-dependent global phase and 
\begin{equation}
\label{alpha_app}
\alpha_k = \frac{J_k(-\xi)}{\kappa/2 - i \left(\Delta_0 + G \bar{x} - k \omega_{\rm m} \right)},
\end{equation}
with $J_k(z)$ being the $k$th Bessel function of the first kind and $\xi = G A / \omega_{\rm m}$ is the dimensionless mechanical modulation strength \cite{aspelmeyer_2014, marquardt_2006, ludwig_2008, krause_2015}. We point out that in this expression for $\alpha_k$, we have explicitly written out the cavity detuning $\Delta = \Delta_0 + G \bar{x}$, as we wish to be more transparent with the $\bar{x}$ term throughout this section for completeness. Note, however, that for the experiment considered in this work, the affect of adding this $G \bar{x}$ term to the bare cavity detuning is negligible, such that $\Delta \approx \Delta_0$. This is demonstrated by the fact that even at the largest optical power input to the device ($P_{\rm in}$ = 139 $\mu$W), we find the maximum static displacement of the resonator to be $\bar{x}_{\rm max}$ = 47 pm, causing a shift in the detuning that is at most $G \bar{x}_{\rm max} = 38$ MHz $= 0.024 \kappa$.

\begin{figure}[h!]
\centerline{\includegraphics[width=\columnwidth]{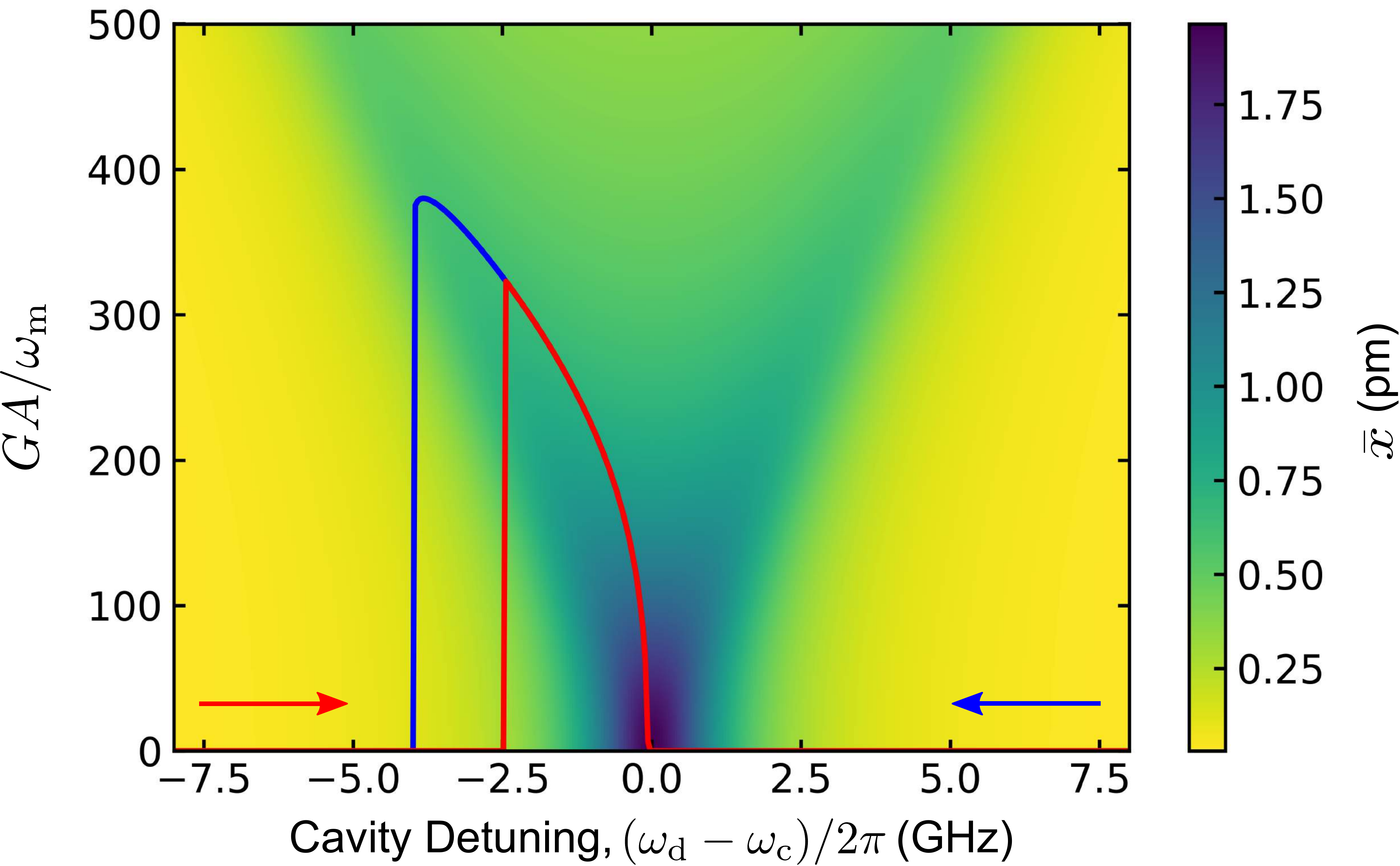}}
\caption{{\label{fig10}} A color plot of $\bar{x}$ versus mechanical amplitude and cavity detuning for the conditions given in Fig.~\ref{fig4}. The solid lines are the mechanical amplitudes that are traced out for detuning sweeps performed in the direction of the corresponding colored arrows (see Fig.~\ref{fig4}). Thus, these contours denote the physical values of $\bar{x}$ that are realized in this situation.}
\end{figure}

We now look to determine the quantities $\bar{x}$, $\delta \omega_{\rm m}$, and $\delta \Gamma_{\rm m}$ in terms of the mechanical amplitude $A$ using this nonlinear optomechanical treatment. Starting by taking the time average of Eq.~\eqref{xeomcl} ({\it i.e.}, balancing the time-averaged forces of the system) we find
\begin{equation}
\label{xbarnl}
\bar{x}(A) = \frac{\hbar G \kappa_{\rm e} |\bar{a}_{\rm in}|^2}{m \omega_{\rm m}^2} \left( 1 + \beta \right) \sum_{k=-\infty}^\infty |\alpha_k|^2,
\end{equation}
where we have used the fact that $\displaystyle \braket{\ddot{x}(t)} = \braket{\dot{x}(t)} = 0$ and $\braket{x(t)} = \bar{x}$, as well as \cite{marquardt_2006, ludwig_2008, metzger_2008b}
\begin{gather}
\label{tavea2}
\braket{|a(t)|^2} = \kappa_{\rm e} |\bar{a}_{\rm in}|^2 \displaystyle \sum_{k=-\infty}^\infty |\alpha_k|^2, \\
\label{taveinta2}
\braket{\frac{\beta}{\tau} \int_{-\infty}^t e^{-\frac{t-t'}{\tau}} |a(t')|^2 dt'} = \beta \kappa_{\rm e} |\bar{a}_{\rm in}|^2 \displaystyle \sum_{k=-\infty}^\infty |\alpha_k|^2. 
\end{gather}
Note that since $\alpha_k$ is implicitly dependent on $\bar{x}$, Eq.~\eqref{xbarnl} represents a transcendental equation for $\bar{x}$ in terms of $A$ and $\Delta_0$ (see Fig.~\ref{fig10}), which in general must be solved numerically. 

Next, we multiply Eq.~\eqref{xeomcl} by $\dot{x}(t)$ and again take the time average, balancing the time-averaged power of the system. In this case, we find
\begin{equation}
\label{dGmnl_app}
\delta \Gamma_{\rm m}(A) = \frac{2 \hbar G \kappa_{\rm e} |\bar{a}_{\rm in}|^2}{A m \omega_{\rm m}} \sum_{k=-\infty}^\infty {\rm Im} \left\{ \alpha_k \alpha_{k+1}^* \left( 1 + \frac{\beta}{1 - i \omega_{\rm m} \tau} \right) \right\},
\end{equation}
where we have now used $\braket{\ddot{x}(t) \dot{x}(t)} = \braket{x(t) \dot{x}(t)} = 0$ and $\braket{\dot{x}^2(t)} = \omega_{\rm m}^2 A^2 /2$, along with \cite{marquardt_2006, ludwig_2008, metzger_2008b}
\begin{gather}
\label{tavea2xdot}
\braket{|a(t)|^2 \dot{x}(t)} = -A \omega_{\rm m} \kappa_{\rm e} |\bar{a}_{\rm in}|^2 \sum_{k=-\infty}^\infty {\rm Im} \left\{ \alpha_k \alpha_{k+1}^* \right\}, \\
\label{taveinta2xdot}
\begin{aligned}
&\braket{\int_{-\infty}^t e^{-\frac{t-t'}{\tau}} |a(t')|^2 \dot{x}(t) dt'} \\
&= -A \omega_{\rm m} \kappa_{\rm e} |\bar{a}_{\rm in}|^2 \displaystyle \sum_{k=-\infty}^\infty {\rm Im} \left\{ \frac{ \beta \alpha_k \alpha_{k+1}^*}{1 - i \omega_{\rm m} t} \right\},
\end{aligned}
\end{gather}
as well as the fact that $\delta \Gamma_{\rm m}(A) = - \Gamma_{\rm m}$ in the nonlinear regime. Finally, it can also be shown (see Ref.~\cite{ludwig_thesis} for example) that the spring effect in the nonlinear optomechanical regime will be given by
\begin{equation}
\label{dwmnl_app}
\delta \omega_{\rm m}(A) = -\frac{\hbar G \kappa_{\rm e} |\bar{a}_{\rm in}|^2}{A m \omega_{\rm m}} \sum_{k=-\infty}^\infty {\rm Re} \left\{ \alpha_k \alpha_{k+1}^* \left( 1 + \frac{\beta}{1 - i \omega_{\rm m} \tau} \right) \right\}.
\end{equation}
Interestingly, one can use the time-averaged energy balance equation [by multiplying Eq.~\eqref{xeomcl} by $x(t)$ and time-averaging], along with the relations $\braket{\dot{x}(t) x(t)} = 0$, $\braket{\ddot{x}(t) x(t)} = -\omega_{\rm m}^2 A^2 / 2$, and $\braket{x^2(t)} = \bar{x}^2 + A^2/2$, to show that $\delta \omega_{\rm m}(A) = 0$ while $A$ is large enough that the system remains in the nonlinear regime. This frequency locking effect, coupled with the reduction of the resonance linewidth, is indicative of phonon lasing in the mechanical resonator \cite{aspelmeyer_2014}.

We are also interested in how mechanical self-oscillations affect the transmission of the optical field through the cavity. To do this, we consider the optical field output from the cavity, which can be found using input-output theory as $a_{\rm out}(t) = a_{\rm in}(t) - \sqrt{\kappa_{\rm e}} a(t)$ \cite{aspelmeyer_2014}. Inserting Eq.~\eqref{aoft_app} into this expression, while only considering the time-independent terms, we find the amplitude-dependent DC transmission through the cavity as
\begin{equation}
\label{Tdcnl_app}
\mathcal{T}_{\rm DC}(A) = \frac{\bar{a}_{\rm out}}{\bar{a}_{\rm in}} = 1 - 2 \kappa_{\rm e} {\rm Re} \left\{ \sum_{k=-\infty}^\infty J_{-k}(\xi) \alpha_k \right\} + \kappa_{\rm e}^2 \sum_{k=-\infty}^\infty |\alpha_k|^2.
\end{equation}
We conclude this section by noting that in the regime of small mechanical oscillations ({\it i.e.}, $\xi \ll 1$), each of the amplitude-dependent quantities above approach their linearized counterpart. That is, Eq.~\eqref{xbarnl} $\rightarrow$ Eq.~\eqref{xbar}, Eq.~\eqref{dGmnl_app} $\rightarrow$ Eq.~\eqref{dGm} and Eq.~\eqref{dwmnl_app} $\rightarrow$ Eq.~\eqref{dwm}, while Eq.~\eqref{Tdcnl_app} approaches its linearized version given by \cite{aspelmeyer_2014}
\begin{equation}
\label{Tdclin}
\mathcal{T}_{\rm DC}^{\rm lin} = 1 - \kappa_{\rm e} (\kappa - \kappa_{\rm e}) |\chi_{\rm c}(0)|^2 = 1 - \frac{\kappa_{\rm e} \kappa_{\rm i}}{\Delta^2 + (\kappa / 2 )^2}.
\end{equation}

\subsection{Integral Approximations}
\label{intapp}

\begin{figure}[h!]
\centerline{\includegraphics[width=3.2in]{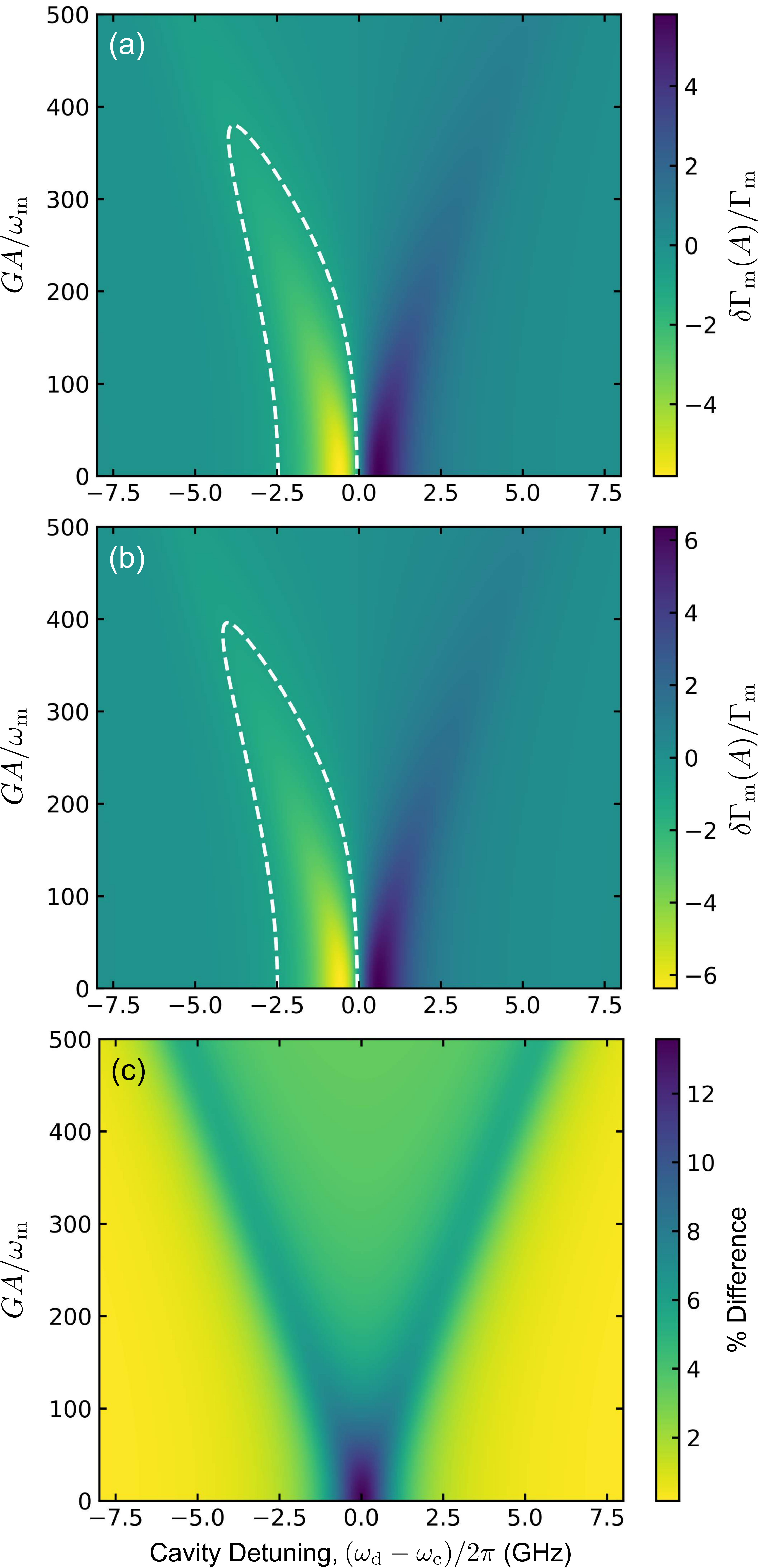}}
\caption{{\label{fig11}} Attractor diagrams of $\delta \Gamma_{\rm m}(A) / \Gamma_{\rm m}$ (color scale) for the device parameters given in Fig.~\ref{fig4} using (a) the exact sum formalism of Eqs.~\eqref{xbarnl} and \eqref{dGmnl_app}, and (b) the integral approximations given by Eqs.~\eqref{xbarint} and \eqref{dGmint}. The white dashed line indicates the contour of $\delta \Gamma_{\rm m}(A) / \Gamma_{\rm m} = -1$, demarcating the region of self-oscillations. For the sums in (a), terms up to $k = \pm 1000$ were used, while the integrals in (b) were performed using a numerical solver (trapezoidal method). In (c), we show the percent difference between the attractor diagrams given in (a) and (b). Here, we highlight the fact that over the displayed detuning and amplitude range, there is at most a 13.6\% difference between the sum and integral methods for calculating $\delta \Gamma_{\rm m}(A)$, which is located near zero detuning for small mechanical amplitudes.}
\end{figure}

Though the expressions in the previous section provide an exact representation for the optomechanical shift in mechanical equilibrium position, damping and spring effects, computing these quantities numerically can be cumbersome. This is due to the fact that in order to accurately model the nonlinear behavior of the optomechanical system, the number of terms that one must keep for each of the sums found in Eqs.~\eqref{xbarnl}--\eqref{dwmnl_app} is on the order of $\xi$, which is in the range of 100 to 1000 for the device studied here [see Fig.~\ref{fig4}(e)]. Fortunately, it was shown by Metzger {\it et al.}~\cite{metzger_2008b} that in the non-sideband-resolved regime, the integral in Eq.~\eqref{Fptcl} can be performed directly by assuming the optical intensity inside the cavity adiabatically follows the quasistatic motion of the mechanical resonator. This allows for a simpler, more computationally efficient treatment of the nonlinear optomechanical system considered in this work, with minimal error in the final results when compared to those given in Eqs.~\eqref{xbarnl}, \eqref{dGmnl_app}, and \eqref{dwmnl_app} (see Fig.~\ref{fig11} for instance). Here we provide a brief overview of this method, resulting in approximate expressions for each of the nonlinear optomechanical properties given in the previous section.

As mentioned above, for this integral approach we immediately assume the non-sideband-resolved regime, such that the optical field in the cavity reacts nearly instantaneously to the resonator's mechanical motion \cite{metzger_2008b}. In this case, we treat $x(t)$ as a quasistatic variable, inputting the ansatz given by Eq.~\eqref{xoft} into Eq.~\eqref{aeomcl}, allowing us to directly solve for the cavity field amplitude as
\begin{equation}
\label{aquasi}
a(t) = \frac{\sqrt{\kappa_{\rm e}} \bar{a}_{\rm in}}{\kappa / 2 - i \left[ \Delta + G\bar{x} + G A \cos(\omega_{\rm m} t) \right]}.
\end{equation}
Using this approximate expression for the cavity field amplitude, we can again take the time average of Eq.~\eqref{xeomcl} to find an integral form for $\bar{x}$ as
\begin{equation}
\label{xbarint}
\bar{x}(A) = \frac{\hbar G \kappa_{\rm e} |\bar{a}_{\rm in}|^2 (1 + \beta)}{2 \pi m \omega_{\rm m}^2} \int_0^{2 \pi} \frac{d \phi}{(\kappa / 2)^2 + (\Delta + G \bar{x} + G A \cos \phi)^2}.
\end{equation}
In comparing this expression with what was found for $\bar{x}$ in Appendix \ref{nlom}, we find that this approximation is equivalent to replacing the sum in Eq.~\eqref{xbarnl} with an integral according to
\begin{equation}
\label{sumintappxb}
\sum_{k=-\infty}^\infty |\alpha_k|^2 \approx \frac{1}{2 \pi} \int_0^{2 \pi} \frac{d \phi}{(\kappa / 2)^2 + (\Delta + G \bar{x} + G A \cos \phi)^2}.
\end{equation}
Furthermore, this integral can be solved analytically, resulting in
\begin{equation}
\begin{split}
\label{xbarintana}
&\int_0^{2 \pi} \frac{d \phi}{(\kappa / 2)^2 + (\Delta + G \bar{x} + G A \cos \phi)^2} \\ 
&= \frac{2 \pi \sqrt{2}}{\kappa \sqrt{\mathcal{A}(\Delta)}} \sqrt{\sqrt{\mathcal{A}(\Delta)} + \mathcal{B}(\Delta)},
\end{split}
\end{equation}
where $\mathcal{A}(\Delta) = \mathcal{B}^2(\Delta) + \kappa^2 (\Delta + G \bar{x})^2$ and $\mathcal{B}(\Delta) = G^2 A^2 + \kappa^2/4 - (\Delta + G \bar{x})^2$. Therefore, we can write $\bar{x}$ in the purely analytical form
\begin{equation}
\label{xbarana}
\bar{x}(A) = \frac{\hbar G \kappa_{\rm e} |\bar{a}_{\rm in}|^2 (1 + \beta) \sqrt{2}}{m \omega_{\rm m}^2 \kappa} \frac{\sqrt{\sqrt{\mathcal{A}(\Delta)} + \mathcal{B}(\Delta)}}{\sqrt{\mathcal{A}(\Delta)}}.
\end{equation}

Performing a similar analysis to determine the integral form for $\delta \Gamma_{\rm m}$, we multiply Eq.~\eqref{xeomcl} by $\dot{x}(t)$ and take the time average, while using the integral approximation for $a(t)$ given by Eq.~\eqref{aquasi} to obtain
\begin{equation}
\label{dGmint}
\begin{split}
&\delta \Gamma_{\rm m} (A) = \frac{\hbar G \kappa_{\rm e} |\bar{a}_{\rm in}|^2 \beta}{ \pi A m \omega_{\rm m}} \frac{\omega_{\rm m} \tau}{\omega_{\rm m}^2 \tau^2 + 1} \\
& \times \int_0^{2 \pi} \frac{\cos \phi d \phi}{(\kappa / 2)^2 + (\Delta + G \bar{x} + G A \cos \phi)^2}.
\end{split}
\end{equation}
Similar to the integral expression for $\bar{x}$, we find that Eq.~\eqref{dGmint} approximates Eq.~\eqref{dGmnl_app} by replacing its sum with the integral
\begin{equation}
\begin{split}
\label{sumintappdGm}
&\sum_{k=-\infty}^\infty {\rm Im} \left\{ \alpha_k \alpha_{k+1}^* \left( 1 + \frac{\beta}{1 - i \omega_{\rm m} \tau} \right) \right\} \\ 
&\approx \frac{\beta}{ 2 \pi} \frac{\omega_{\rm m} \tau}{\omega_{\rm m}^2 \tau^2 + 1} \int_0^{2 \pi} \frac{\cos \phi d \phi}{(\kappa / 2)^2 + (\Delta + G \bar{x} + G A \cos \phi)^2}.
\end{split}
\end{equation}
This integral also has an analytical expression given by
\begin{equation}
\begin{split}
\label{dGmintana}
&\int_0^{2 \pi} \frac{\cos \phi d \phi}{(\kappa / 2)^2 + (\Delta + G \bar{x} + G A \cos \phi)^2} \\
&= -{\rm sgn}(\Delta + G \bar{x}) \frac{ 2 \pi \sqrt{2}}{G A \kappa \sqrt{\mathcal{A}(\Delta)}} \bigg( |\Delta + G \bar{x} | \sqrt{\sqrt{\mathcal{A}(\Delta)} - \mathcal{B}(\Delta)} \\
&- \frac{\kappa}{2} \sqrt{\sqrt{\mathcal{A}(\Delta)} - \mathcal{B}(\Delta)} \bigg),
\end{split}
\end{equation}
where ${\rm sgn}(z)$ is the signum function. Using this relation, we can then express $\delta \Gamma_{\rm m}$ in the analytical form
\begin{equation}
\begin{split}
\label{dGmana}
&\delta \Gamma_{\rm m} (A) = \frac{- 2\sqrt{2} \hbar \kappa_{\rm e} |\bar{a}_{\rm in}|^2 \beta}{A^2 m \omega_{\rm m} \kappa} \frac{\omega_{\rm m} \tau}{\omega_{\rm m}^2 \tau^2 + 1} \frac{{\rm sgn}(\Delta + G \bar{x})}{\sqrt{\mathcal{A}(\Delta)}} \\ 
&\times \left( |\Delta + G \bar{x} | \sqrt{\sqrt{\mathcal{A}(\Delta)} - \mathcal{B}(\Delta)} - \frac{\kappa}{2} \sqrt{\sqrt{\mathcal{A}(\Delta)} - \mathcal{B}(\Delta)} \right).
\end{split}
\end{equation}
It is also possible to arrive at an integral expression for $\delta \omega_{\rm m}$, which looks similar to Eq.~\eqref{dGmint} except the factor of $\cos(\phi)$ in the numerator of the integrand is replaced with $\sin(\phi)$. This, however, results in an integral that evaluates to zero, as one would expect in the self-oscillating regime (see Appendix \ref{nlom}), and therefore offers no new insight into Eq.~\eqref{dwmnl_app}.

In Fig.~\ref{fig11}, we compare the attractor diagrams of $\delta \Gamma_{\rm m}$ for the optomechanical device studied in this work generated using both the exact sums given in Eqs.~\eqref{xbarnl} and \eqref{dGmnl_app}, as well as the integral approximations of Eqs.~\eqref{xbarint} and \eqref{dGmint}. As our device exists deeply in the non-sideband-resolved regime ($\kappa / \omega_{\rm m} \approx 180$), the integral approximations presented in this section accurately model its nonlinear optomechanical behavior. This is demonstrated by the fact the percent difference in $\delta \Gamma_{\rm m}$ between these two methods is at most 13.6\% for the conditions given in Fig.~\ref{fig4} [see Fig.~\ref{fig11}(c)]. Furthermore, we note that while the integral approach slightly overshoots the value of $\delta \Gamma_{\rm m}$, it still provides an excellent approximation of the mechanical amplitude, as can be seen by the nearly matching contour lines in Figs.~\ref{fig11}(a)/(b).

We conclude this section by noting that while we have used Eqs.~\eqref{xbarint}--\eqref{dGmana} for preliminary assessment of our optomechanical device, as well as the computationally intensive calculations associated with the varying power measurements shown in Fig.~\ref{fig5}, the fits and attractor diagram in Fig.~\ref{fig4} were determined using the exact expressions given by Eqs.~\eqref{xbarnl}--\eqref{dwmnl_app}.

\subsection{Optical Power Transfer}
\label{optpow}

\begin{figure}[h!]
\centerline{\includegraphics[width=2.5in]{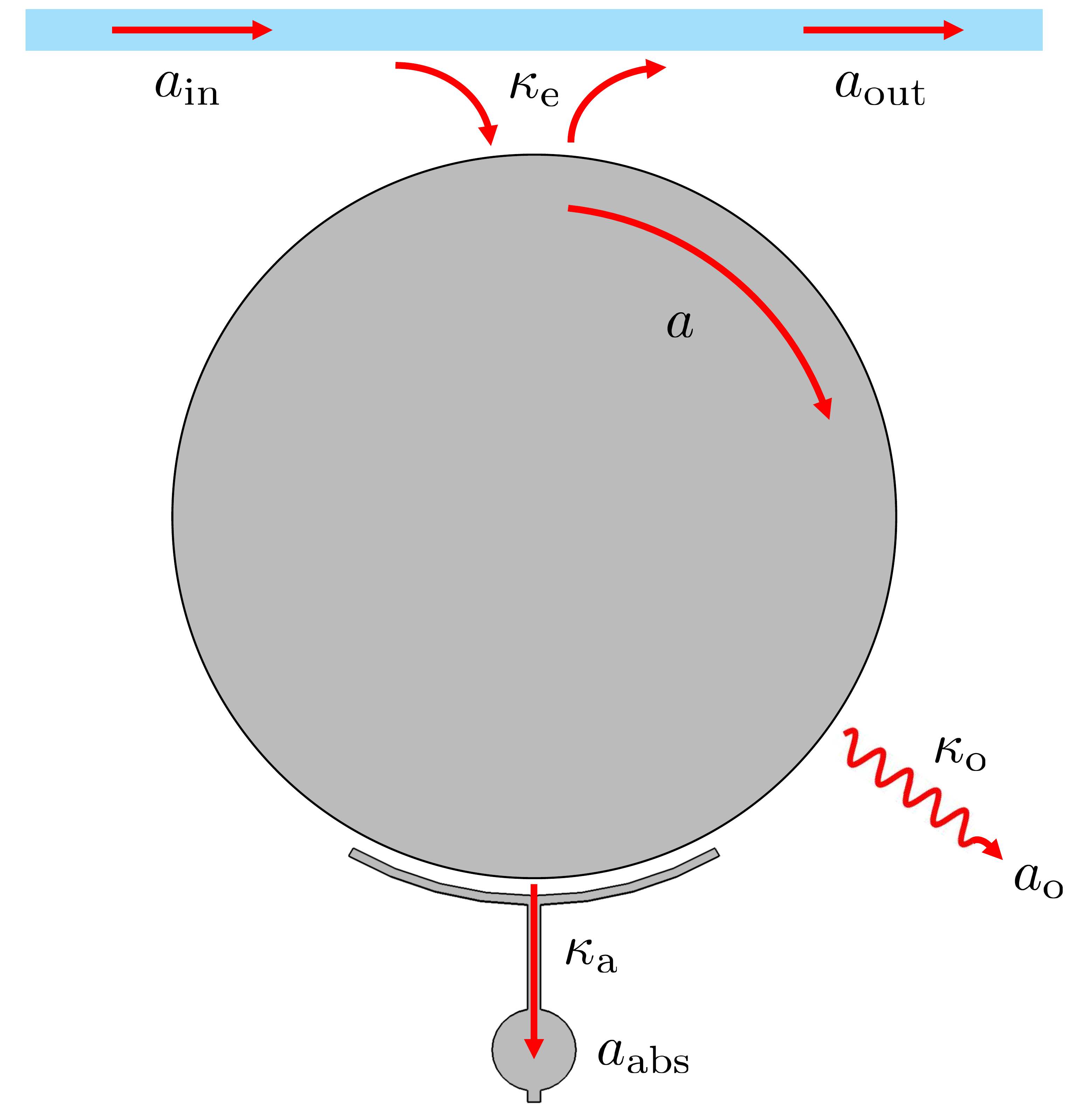}}
\caption{{\label{fig12}} Schematic illustrating the flow of the optical field through the optomechanical cavity.}
\end{figure}

We conclude this section by using the input-output formalism that was introduced previously in Appendix \ref{nlom}, along with the conservation of energy, to investigate the power input to and output from the optical cavity, as well as that absorbed by the mechanical resonator and dissipated via other loss mechanisms (see Fig.~\ref{fig12}). Restricting ourselves to a linearized classical treatment (all quantum effects will average out to zero), we begin with the power input to the optical cavity, which can be expressed in terms of the field amplitude input by the external waveguide as $P_{\rm in} = \hbar \omega_{\rm d} |\bar{a}_{\rm in}|^2$ \cite{aspelmeyer_2014}. The power recollected by this external waveguide (and subsequently sent to our detection apparatus) is then simply given by (see Appendix \ref{nlom})
\begin{equation}
\label{Pout}
P_{\rm out} = \hbar \omega_{\rm d} |\bar{a}_{\rm out}|^2 = P_{\rm in} \left[ 1 - \frac{\kappa_{\rm e} \kappa_{\rm i}}{\Delta^2 + (\kappa / 2 )^2}  \right].
\end{equation}
Meanwhile, the power absorbed by the mechanical resonator will be given by
\begin{equation}
\label{Pabs}
P_{\rm abs} = \hbar \omega_{\rm d} |\bar{a}_{\rm abs}|^2 = P_{\rm in} \frac{\kappa_{\rm e} \kappa_{\rm a}}{\Delta^2 + (\kappa / 2 )^2}.
\end{equation}
Finally, power lost to other cavity dissipation channels is found as
\begin{equation}
\label{Po}
P_{\rm o} = \hbar \omega_{\rm d} |\bar{a}_{\rm o}|^2 = P_{\rm in} \frac{\kappa_{\rm e} \kappa_{\rm o}}{\Delta^2 + (\kappa / 2 )^2},
\end{equation}
where $\bar{a}_{\rm o}$ is the steady-state field amplitude associated with these damping mechanisms. We note that Eqs.~\eqref{Pout}--\eqref{Po} obey the conservation of energy in the sense that $P_{\rm in} = P_{\rm out} + P_{\rm abs} + P_{\rm o}$.

Equation \eqref{Pabs} has very important consequences for optically induced heating of the mechanical resonator. This is due to the fact that even if the same amount power is input to the cavity, differing values of $\kappa_{\rm e}$ and $\kappa_{\rm i}$, and therefore $\kappa$, can cause varying amounts of power to be absorbed by the resonator, causing it to heat to different temperatures. For instance, inputting the values of $\kappa$ and $\kappa_{\rm e}$ from the two different coupling conditions for the data found in Figs.~\ref{fig4} and \ref{fig5} (while assuming $\kappa_{\rm a}$ remains the same in each case), we find the power absorbed by the resonator (on cavity resonance) to be approximately 25\% larger for the data in Fig.~\ref{fig5} compared to Fig.~\ref{fig4}. This, coupled with the rapid increase in $\beta$ at low optical input powers, likely accounts for the fact that $\beta$ differs between these two data sets for similar input powers.

\section{Radiation-Pressure-Driven Attractor}
\label{radatt}

\begin{figure}[h!]
\centerline{\includegraphics[width=\columnwidth]{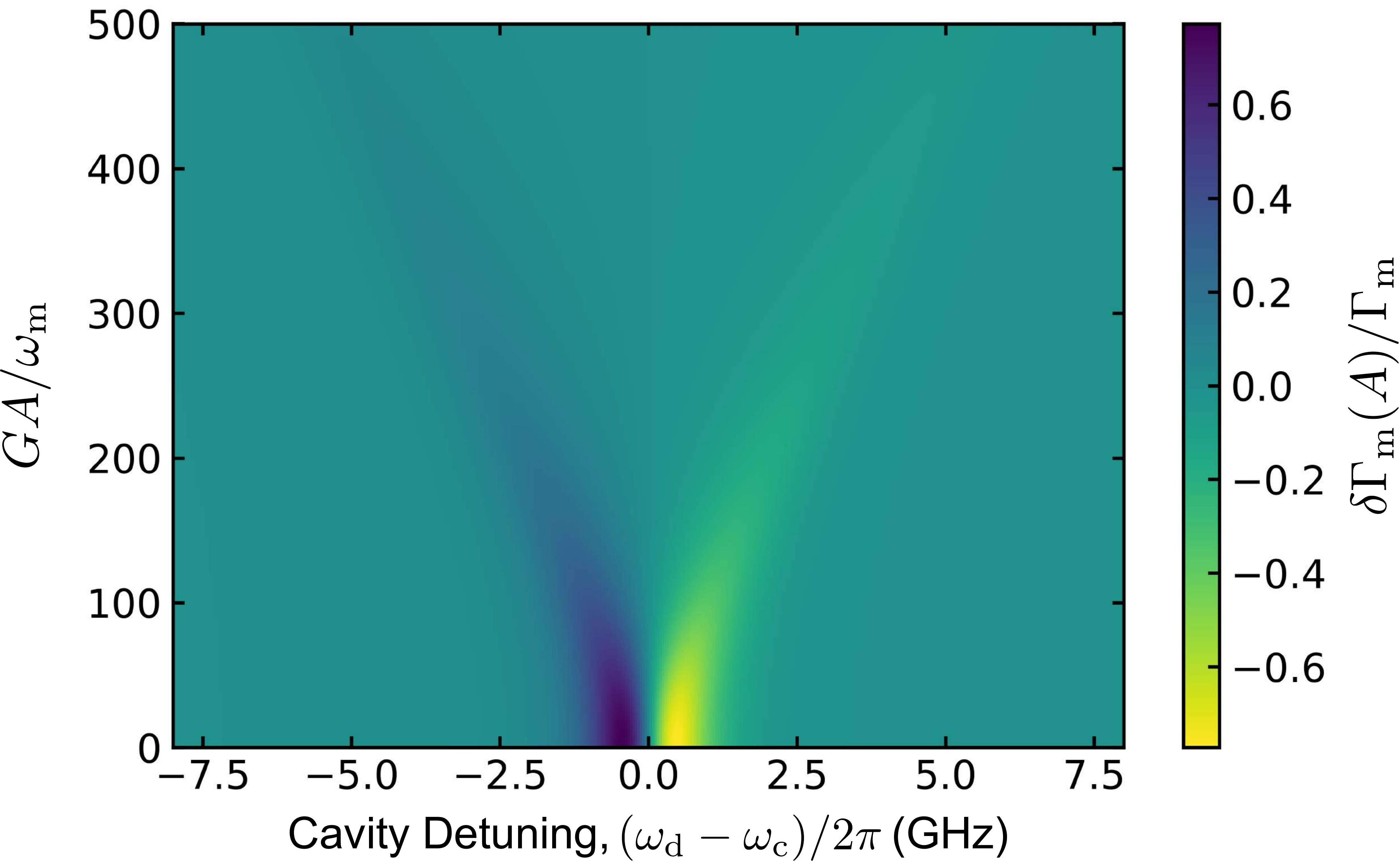}}
\caption{{\label{fig13}} Attractor diagram of $\delta \Gamma_{\rm m}(A) / \Gamma_{\rm m}$ for the device parameters given in Fig.~\ref{fig4}, except with $\beta = 0$ such that only radiation pressure effects are present. We note that not only does the sign of the optomechanical damping reverse, restoring what one would expect for a radiation-pressure-driven system, but the damping force is no longer strong enough to induce mechanical self-oscillations at this power ($P_{\rm in}$ = 10.1 $\mu$W), as demonstrated by the fact that $\delta \Gamma_{\rm m}(A) / \Gamma_{\rm m} > -1$ for all cavity detunings.}
\end{figure}

In order to understand just how substantial photothermal effects are in determining the optomechanical properties of the studied device, it is interesting to investigate the attractor diagram with only the radiation pressure force present. To do this, we have produced an attractor diagram for the device parameters given in Fig.~\ref{fig4} while setting $\beta = 0$ such that the photothermal force is negated. The result is drastic (see Fig.~\ref{fig13}), as the absence of photothermal effects causes the detuning dependence of the optomechanical damping to reverse. Furthermore, we show that with the radiation pressure force alone, this system is no longer able to be driven into self-oscillations at the considered optical input power of $P_{\rm in} = 10.1$ $\mu$W. It is therefore clear that the addition of photothermal force has significant effects on the optomechanical properties of the system.

\section{Power Dependence of Optomechanical Properties}
\label{powdep}

\subsection{Spring Effect and Damping}
\label{powdepwmGm}

\begin{figure}[h!]
\centerline{\includegraphics[width=3in]{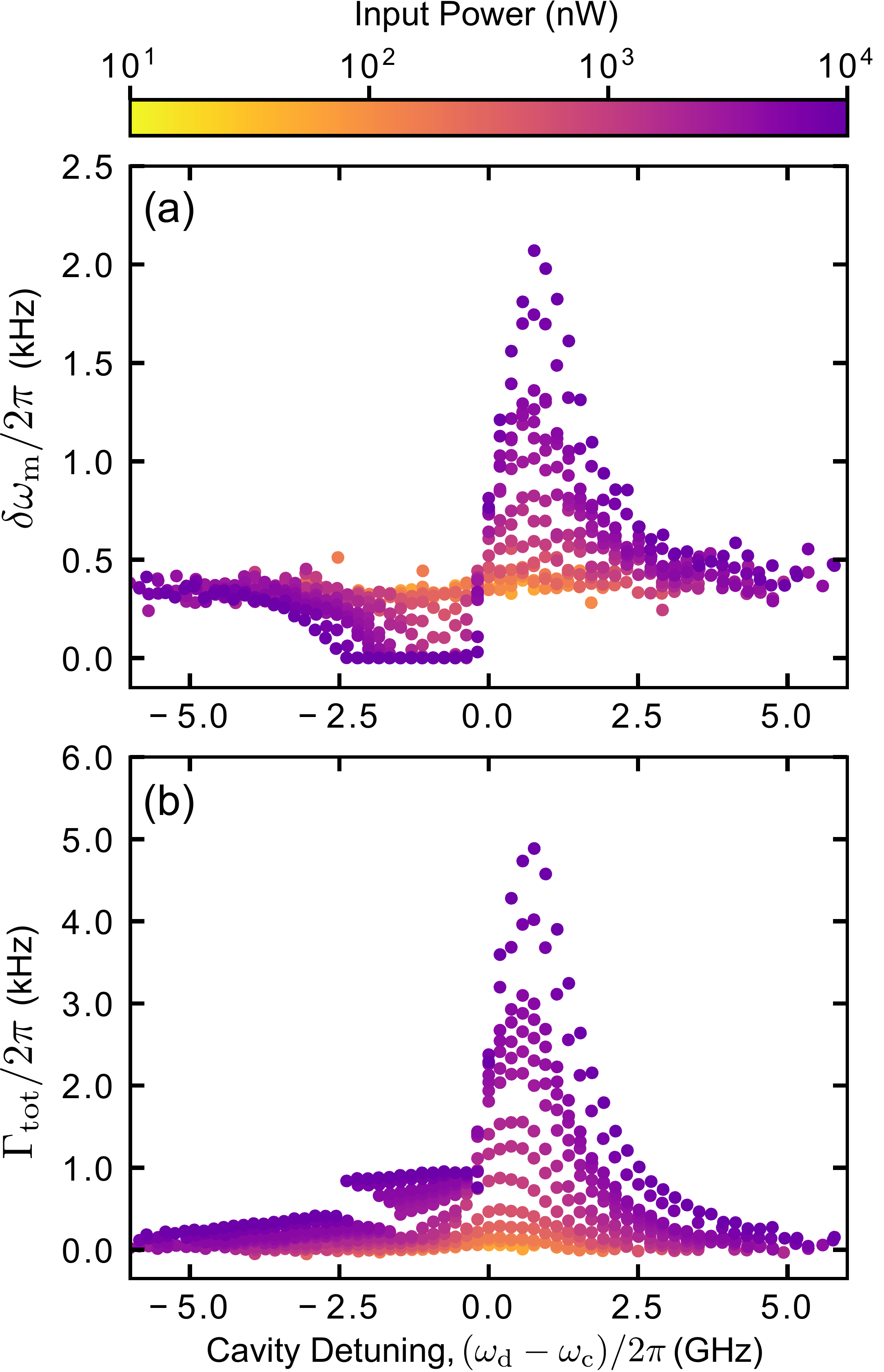}}
\caption{{\label{fig14}} Optomechanical (a) spring effect and (b) damping over three orders of magnitude in input optical power, ranging from 10 nW to 10 $\mu$W. Even at very low powers, optomechanical damping occurs for a blue-detuned optical drive, qualitatively matching the detuning dependence of the optomechanical spring effect.}
\end{figure}

Along with the studies shown in Figs.~\ref{fig3}--\ref{fig5}, we have also investigated the optomechanical properties of our device over three orders of magnitude in input optical power from 10 nW to 10 $\mu$W, as seen in Fig.~\ref{fig14}. Most importantly, we observe that for all input powers, the optomechanical damping exhibits the same qualitative behavior as the optical spring effect, similar to what is seen in Figs.~\ref{fig3}(d)/(f) and \ref{fig4}(c)/(d). Such a power-dependence is in agreement with an optomechanical damping caused by dueling radiation pressure and photothermal forces, as both of these effects scale with an identical power-dependence, as can be seen in Eqs.~\eqref{dGm} and \eqref{dGmnl_app}.

\subsection{Hysteresis in Optical Transmission}
\label{opthyst}

\begin{figure}[h!]
\centerline{\includegraphics[width=3in]{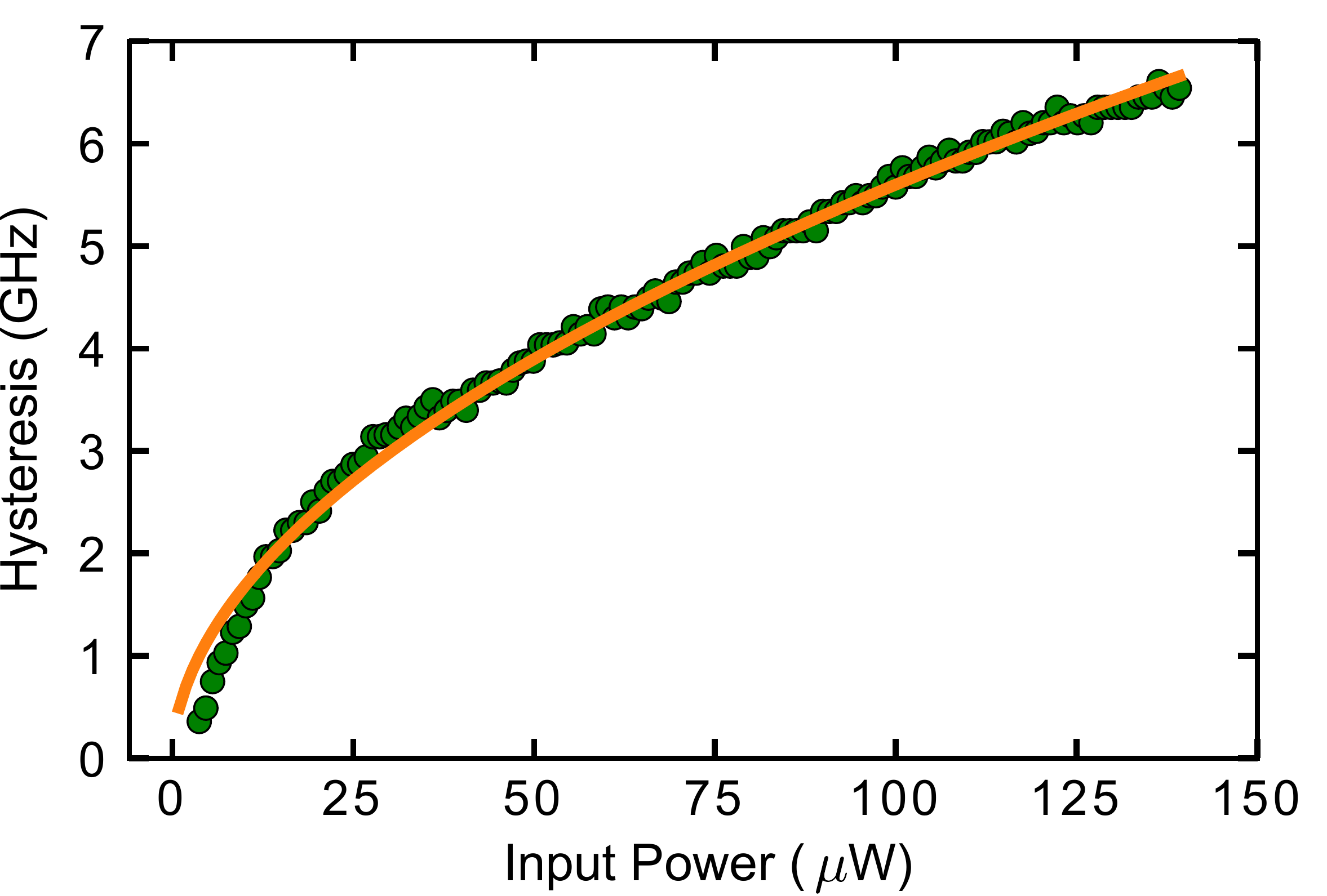}}
\caption{{\label{fig15}} Hysteresis spacing of the DC transmission through the optical cavity versus power (green circles) extracted from the data in Figs.~\ref{fig5}(a)/(c). The orange line is a fit to a power law, from which we extract an exponent of 0.52, indicating a near square-root dependence of the hysteresis spacing on input optical power.}
\end{figure}

Due to the bistable nature of the attractor diagram shown in Fig.~\ref{fig4}(e), amplification of the mechanical resonator's motion results in hysteretic behavior of the DC transmission through the cavity depending on whether the pump beam swept from its red or blue side. As the optical power input to the cavity is increased, optomechanical amplification occurs for a larger number of pump detunings, causing this hysteresis spacing to expand. In Fig.~\ref{fig15}, we showcase this effect for the data in Figs. 5(a)/(c), where the we demonstrate that at high input powers ($P_{\rm in} \gtrsim$ 25 $\mu$W), the hysteresis spacing roughly obeys a square-root dependence on power.

\section{Photothermal Relaxation Time}
\label{tauthsec}

The photothermal time constant $\tau$ that was introduced in Eq.~\eqref{Fpt} is a very important quantity that sets the time scale, and in some instances the strength, of photothermally-driven optomechanical effects. For a thin beam of rectangular cross section, Zener showed that this time constant is dominated by thermal relaxation of the fundamental mode of the beam \cite{zener_1948, lifshitz_2000}, resulting in
\begin{equation}
\label{tauth}
\tau = \frac{l^2 C_p}{\pi^2 k_{\rm th}},
\end{equation}
where $l$, $\rho$, $C_p$, and $k_{\rm th}$ are the length, density, specific heat capacity (per unit mass) at constant pressure, and thermal conductivity of the beam. For a nonmagnetic, crystalline insulator ({\it i.e.}, silicon), the thermal properties of the material are governed by its phonons, such that at low temperatures the heat capacity can be determined according to the Debye model as \cite{pobell_2007}
\begin{equation}
\label{heatcap}
C_p = \frac{2 \pi^5 k_{\rm B}^4 T^3 }{5 \hbar^3 c_{\rm s}^3},
\end{equation}
where $c_{\rm s}$ is the effective speed of sound of the phonons. For silicon, this effective speed is given by $\displaystyle c_{\rm s} = \sqrt[3]{ \frac{1}{3} \sum_i \frac{1}{c_i^3}}$ = 5718 m/s, where the sum is over the single longitudinal and two transverse polarizations, each with a speed of sound given by $c_l =$ 9148 m/s, $c_{t_1} =$ 4679 m/s, and $c_{t_2} =$ 5857 m/s \cite{hauer_2018}. Furthermore, treating the phonons as a diffuse, noninteracting gas, we can express the thermal conductivity as \cite{pobell_2007}
\begin{equation}
\label{kth}
k_{\rm th} = \frac{1}{3} C_p \Lambda c_{\rm s} = \frac{2 \pi^2 k_{\rm B}^4 T^3 \Lambda}{15 \hbar^3 c_{\rm s}^2},
\end{equation}
where $\Lambda$ is the phonon mean free path, which is in general temperature-dependent \cite{weber_1991}. However, as pointed out by Casimir \cite{casimir_1938}, below a certain temperature this mean free path will become comparable to the dimensions of the system, such that it will be limited by the device's finite size. We note that for the resonator studied in this work, this transition temperature is approximately 100 K \cite{weber_1991}, far exceeding its experimental operating temperature. For a beam with a rectangular cross section, this boundary-limited mean free path is given by $\Lambda = 1.12 \sqrt{w d}$, where $w$ and $d$ are the width and thickness of the beam \cite{casimir_1938, ziman_2001, heron_2009, heron_2010, cahill_2014}. Inputting this relation, along with Eqs.~\eqref{heatcap} and \eqref{kth}, into Eq.~\eqref{tauth} allows us to express the thermal relaxation time as
\begin{equation}
\label{tauth2}
\tau = \frac{3 l^2}{1.12 \pi^2 c_{\rm s} \sqrt{w d}}.
\end{equation}
Therefore, at low temperatures and for small geometries ($T \lesssim$ 100 K for dimensions on the order of 100 nm \cite{weber_1991}), the thermal relaxation time in silicon depends only on the geometry and speed of sound of the system, which are to first order temperature-independent.

\begin{figure}[t!]
\centerline{\includegraphics[width=3in]{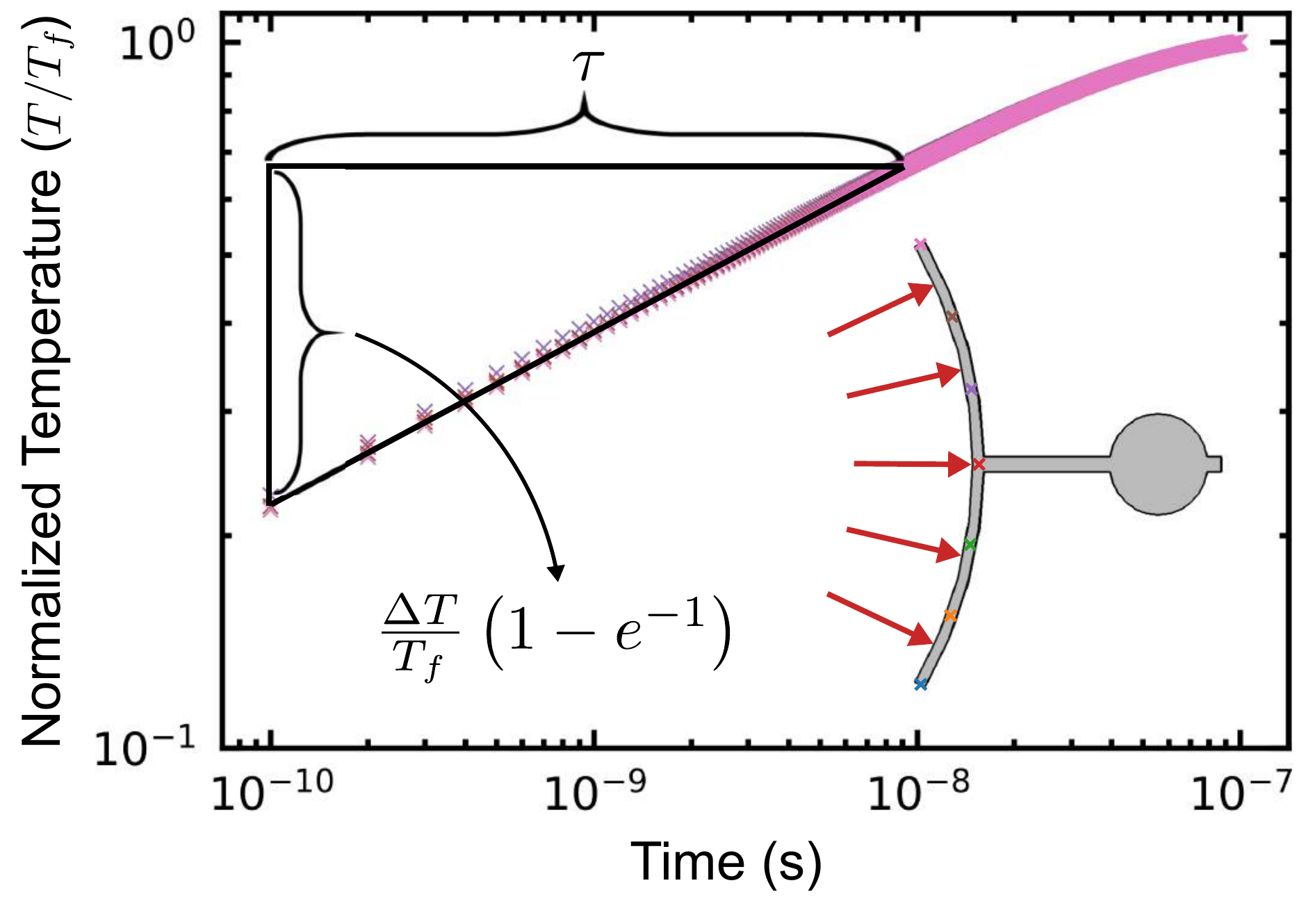}}
\caption{{\label{fig16}} Plot of the normalized temperature $T/T_f$ versus time for the finite element heating simulation used to determine the thermal relaxation time of the resonator. Inset is a schematic illustrating the simulation procedure, whereby a uniform heat load of 6 $\mu$W is applied to the surface of the resonator indicated by the red arrows, while its temperature profile is probed in time at the positions denoted by the colored crosses. By normalizing each of these extracted data sets (color-coded to match the corresponding probe point) to its final temperature, a universal heating trend along the curved portion of the resonator is demonstrated. From these normalized profiles, we determine the thermal relaxation time of the device ($\tau$ = 9.1 ns) as the time required for the temperature to increase from its initial value by an amount $\Delta T \left( 1 - e^{-1} \right)$ [see Eq.~\eqref{Tvtime}].}
\end{figure}

While the above description works well for determining the thermal time constant for the geometry of a simple beam with a uniform rectangular cross section, it is unclear if such an analysis applies to the complex device structure studied here (see Fig.~\ref{fig8}). Therefore, we have performed finite element simulations to accurately determine the thermal relaxation time for this device. Here, the phonon mean free path is limited by the smallest dimension of our resonator ({\it i.e.}, $w_3$ = 151 nm in Fig.~\ref{fig8}), leading to $\Lambda = 1.12 \sqrt{w_3 d}$ = 218 nm, where we have taken $d$ = 250 nm as the thickness of the silicon device. Using this value for the mean free path, along with the temperature-dependent expressions for the specific heat capacity and thermal conductivity found in Eqs.~\eqref{heatcap} and \eqref{kth}, we simulate the heating of the device due to absorption of laser power. To do this, the laser-driven heating was approximated as a uniform heat load applied to the inner surface of the resonator facing the disk [see inset of Fig.~\ref{fig16}]. The magnitude of this heat load is chosen to be $P_{\rm abs}$ = 6 $\mu$W to match the expected absorbed power for the conditions associated with Fig.~\ref{fig4} ({\it i.e.}, $\kappa$ = 2.04 GHz, $\kappa_{\rm e}$ = 0.38 GHz, and $P_{\rm in}$ = 10.1 $\mu$W), while also assuming $\kappa_{\rm a} = \kappa_{\rm i}$ and $\Delta = 0$ (see Appendix \ref{optpow}). The simulated temperature increase of the resonator as a function of time was then monitored at seven equally spaced points along its rounded portion, as shown in Fig.~\ref{fig16}. Continuing with the Zener approximation ({\it i.e.}, the majority of this thermal relaxation occurs through the fundamental mode of the resonator), we expect the temperature at each of these points to increase according to \cite{abdi_2012b, carslaw_1959}
\begin{equation}
\label{Tvtime}
T(t) = T_0 + \Delta T \left( 1 - e^{-t/\tau} \right).
\end{equation}
Here $\Delta T = T_f - T_0$ is the difference between the resonator's temperature $T_0$ at $t=0$ when the heat load is initially applied and its final equilibrium temperature $T_f$ that is reached for $t \gg \tau$. We note that while each point on the resonator heats from an initial temperature of $T_0$ = 4.2 K to varying equilibrium temperatures ranging from $T_f$ = 43 K to $T_f$ = 48 K, when normalized by these final temperatures, each simulated data set collapses onto a single universal trace (see Fig.~\ref{fig16}).  Therefore, we can use Eq.~\eqref{Tvtime} to extract the thermal relaxation time as the average time required for the resonator to heat from $T_0$ to $T_0 + \Delta T (1 - e^{-1})$. Performing this calculation for each of these data sets, we find $\tau$ = 9.1 $\pm$ 0.2 ns, where the uncertainty is given by the standard deviation of this distribution.

To conclude this section, we use this simulated value of $\tau$ to evaluate how well our irregular resonator geometry is approximated as a uniform rectangular beam (with width $w_3$ and thickness $d$). This is done by rearranging Eq.~\eqref{tauth2} to obtain the effective thermal length
\begin{equation}
\label{leff}
l_{\rm eff} = \sqrt{ \frac{1.12 \pi^2 c_{\rm s} \tau \sqrt{w_3 d}}{3}}.
\end{equation}
Using the simulated time constant and the parameters for our device, we find this effective length to be $l_{\rm eff}$ = 6.34 $\mu$m. Comparing this value to the total length of our device, $l_{\rm tot} = l_1 + l_2 + l_3 + 2 R_d$ = 7.06 $\mu$m, as measured from the tip of one end of the rounded portion of the resonator to its anchor point, we find that these two lengths agree very well with each other. Thus, our device is well-approximated as a uniform beam provided we introduce a small reduction in its length by a numerical factor of $l_{\rm eff} / l_{\rm tot} = 0.90$.

\end{appendix}

\end{document}